\documentclass[aps,prb,twocolumn,amsmath,showpacs,amssymb]{revtex4}

\usepackage{amsmath}
\usepackage{graphicx}
\usepackage{amssymb}

\newcommand{\pdag}{{\phantom{\dagger}}}

\begin{document}

\title{Helical nuclear spin order in two-subband quantum wires}

\author{Tobias Meng}
\author{Daniel Loss}
\affiliation{Department of Physics, University of Basel, Klingelbergstrasse 82, CH-4056 Basel, Switzerland}

\begin{abstract}
In quantum wires, the hyperfine coupling between conduction electrons and nuclear spins can lead to a (partial) ordering of both of them at low temperatures. By an interaction-enhanced mechanism, the nuclear spin order, caused by RKKY exchange, acts back onto the electrons and gaps out part of their spectrum. In wires with two subbands characterized by distinct Fermi momenta $k_{F1}$ and $k_{F2}$, the nuclear spins form a superposition of two helices with pitches $\pi/k_{F1}$ and $\pi/k_{F2}$, thus exhibiting a beating pattern. This order results in a reduction of the electronic conductance in two steps upon lowering the temperature.
\end{abstract}

\maketitle

\section{Introduction}
Being a hallmark of topological states of matter as well as a versatile platform for engineering quantum computation devices, helical and quasi-helical electron systems have attracted much interest in recent years. Besides their prominent formation as edge states in Quantum Spin Hall samples,\cite{qshe} (quasi-) helical Luttinger liquids can for instance also be generated by applying a magnetic field to a Rashba spin-orbit coupled quantum wire.\cite{streda_03} When brought into contact with an ordinary superconductor, a helical Luttinger liquid can turn into a topological superconductor with zero energy Majorana bound states at the two ends of the wire.\cite{lutchyn_10, oreg_10, alicea_10} Helical Luttinger liquids are however also interesting for their own sake and have for instance applications as spin filters.\cite{streda_03}

In addition to spin-orbit based proposals, it has been realized that quasi-helical Luttinger liquids may also emerge due to helical magnetic fields. The helical field is thereby fully equivalent to the combination of a homogeneous magnetic field and Rashba spin-orbit interactions.\cite{braunecker_prb_10} The interplay of such helical fields (and more generally oscillating fields, for instance generated by nanomagnets) with the usual spin-orbit coupling furthermore allows to extend the topological phase diagram of Rashba spin-orbit coupled quantum wires to exotic phases beyond Majorana bound states.\cite{klinovaja_12,klinovaja_ladders_13}

As an alternative to an applied helical magnetic field, the intrinsic hyperfine coupling between electrons and nuclear spins can lead to the spontaneous formation of a quasi-helical Luttinger liquid at low temperatures. As has been argued in the case of a single subband quantum wire, the hyperfine coupling results in a Ruderman-Kittel-Kasuya-Yosida (RKKY) interaction that diverges at momentum $2k_F$ due to low energy electron backscattering.\cite{braunecker_prl_09,braunecker_prb_09} This in turn induces a helical order of the nuclear spins, which feeds back to the electrons as a magnetic Overhauser field. The latter gaps out half of the electronic degrees of freedom, thereby turning the quantum wire into a quasi-helical one. Importantly, the electronic gap, strongly enhanced by electron-electron interactions, opens exactly around the Fermi level without any fine tuning.\cite{braunecker_prl_09,braunecker_prb_09} The resulting reduction of the conductance\cite{braunecker_prb_09,meng_loss_conductance} has recently been observed in cleaved edge overgrowth GaAs quantum wires.\cite{scheller_nuclear_spin_experiment_arxiv} We expect such helical (nuclear) spin order to be observable not only in hyperfine coupled quantum wires, but more generally in systems that fall into the class of weakly-coupled, electronically one-dimensional Kondo lattice systems of finite size in the so-called RKKY liquid phase when both the Kondo energy scale $k_B\, T_K$ and the direct exchange between the Kondo lattice spins are smaller than the RKKY interaction (the nuclear spins are one possible example of such a Kondo lattice).\cite{sigrist_review} The helical order of the nuclear spins in quantum wires is furthermore intimately related to a proposed ferromagnetic order of the nuclear spins in two-dimensional electron gases.\cite{2deg_spin_suscept_zak} 

The quasi-helical state resulting from gapping out parts of the electronic spectrum by either a combination of a homogeneous magnetic field and spin-orbit interactions, a helical magnetic field, or a spontaneous ordering of the nuclear spins, can in some ways be understood as a helical Luttinger liquid (the spin of the remaining gapless electronic modes is approximately locked to their direction of motion). Their physics is however even richer than the one of an ideal helical liquid. This has been discussed in terms of the spectral properties and the optical conductivity of the wire.\cite{braunecker_prb_12,schuricht_12} For a better distinction from ideal helical Luttinger liquids, the quasi-helical system resulting from gapping out parts of the electronic spectrum has been dubbed spiral Luttinger liquid or spiral spin density wave state. Helical Luttinger liquids can in fact be understood as a special subclass of spiral Luttinger liquids with charge and spin Luttinger parameters $K_c = 1/K_s$.\cite{braunecker_prb_12,hou_09} The study of spiral Luttinger liquids is consequently of specific interest as new phenomena (beyond helical Luttinger liquid physics) can arise. Since furthermore a considerable number of experiments and theoretical proposals are based on spiral Luttinger liquids mimicking helical Luttinger liquids, the understanding of spiral Luttinger liquids is also of practical relevance.

In this work, we discuss how stable the spontaneous formation of helical nuclear spin order in a quantum wire is to the presence of multiple subbands.  Indeed, multi-subband  quantum wires are characterized by different Fermi momenta $k_{Fi}$ for the different bands. One may thus expect low-energy electron backscattering at any combination $k_{Fi}+k_{Fj}$, and it is not obvious how the nuclear spins order in this case. To analyze their interplay with the electrons, we specifically focus on a quantum wire with two subbands that may either correspond to the lowest two bands of a single wire, or alternatively emerge from two sufficiently coupled parallel quantum wires with a single occupied band each. In the latter case, the two subbands are the symmetric and antisymmetric orbitals shared between the two wires. While the experiment reported in reference [\onlinecite{scheller_nuclear_spin_experiment_arxiv}] involves two closely spaced quantum wires, the observed coupling between the wires is rather small. The experiment should thus be interpreted in terms of (almost) decoupled, single subband wires. We are however confident that this first experimental detection of hyperfine induced nuclear spin order will pave the way additional experiments also addressing the multi subband case discussed in the remainder. After deriving the effective one-dimensional model in Sec.~\ref{sec:model}, we list and analyze the possible orders of the nuclear spins in Sec.~\ref{sec:helix_order}. We then discuss that the nuclear spins form two superimposed helices at momenta $2k_{F1}$ and $2k_{F2}$, but not at $k_{F1}+k_{F2}$, and analyze the resulting low energy theory in Sec.~\ref{sec:feedback} in a self-consistent fashion. In Secs.~\ref{sec:order_and_finite_t} and \ref{sec:inter_intra}, we finally turn to the onset and stability of this order.

\section{The model}\label{sec:model}
For concreteness, we specialize to a model of a single quantum wire defined in a two-dimensional GaAs electron gas by virtue of electrostatic gates. The chemical potential, tunable by an electrostatic gate, is chosen such that only the lowest two subbands of the wire are (partially) filled. Nevertheless, our findings remain qualitatively valid for Carbon nanotube samples or the above mentioned parallel wires.

As a central ingredient for our discussion, the electrons interact via screened Coulomb interaction. The screening may for instance be due to mirror charges in the gate electrodes of the sample and is characterized by a screening length larger than the typical length scale associated with the transversal confinement. We neglect the weak spin-orbit interaction (note that the spin orbit length is much larger than typical Fermi wavelengths in GaAs wires and Carbon nanotubes).\cite{charis_gaas_soi_gap_10,kuemmeth_cnt_so_int_exp_09} Expressing the electron Hamiltonian in terms of right-moving and left-moving particles in the two subbands, the Coulomb interaction gives rise to various matrix elements for electrons close to the four Fermi points. If there were no nuclear spins, the electrons could be described as two spinful and gapless Luttinger liquids whose velocities and Luttinger liquid parameters are renormalized by a number of density-density type interactions.\cite{varma_zawadowski_85,meng_11} For simplicity, we will, however, only keep track of charge density interactions since the latter are typically larger than all other electron-electron couplings, resulting in strongly renormalized charge velocities and fairly unrenormalized spin velocities.\cite{cnt_luttinger_param_theory, auslaender_spin_charge_seperation_exp_05, jompol_kc_ks, steinberg_charge_fractionalization_08, macdonald_lutt_param_01} We will come back to the small effect of spin density interactions on a spin ordered state in Sec.~\ref{sec:order_and_finite_t}.

The nuclear spins in the wire are modeled by a Kondo lattice type Hamiltonian. They interact with each other via dipolar and quadrupolar interactions. Most importantly, they are also subject to a hyperfine coupling to the electrons. As will be discussed below, the hyperfine coupling gives rise to an RKKY interaction between the nuclear spins.\cite{simon_loss_07} Similar to single band wires,\cite{braunecker_prb_09} this RKKY interaction overrules any direct coupling between the nuclear spins by orders of magnitude, and we will thus neglect the latter in the remainder.

\begin{figure}
\centering
$(a)$\quad\includegraphics[scale=0.4]{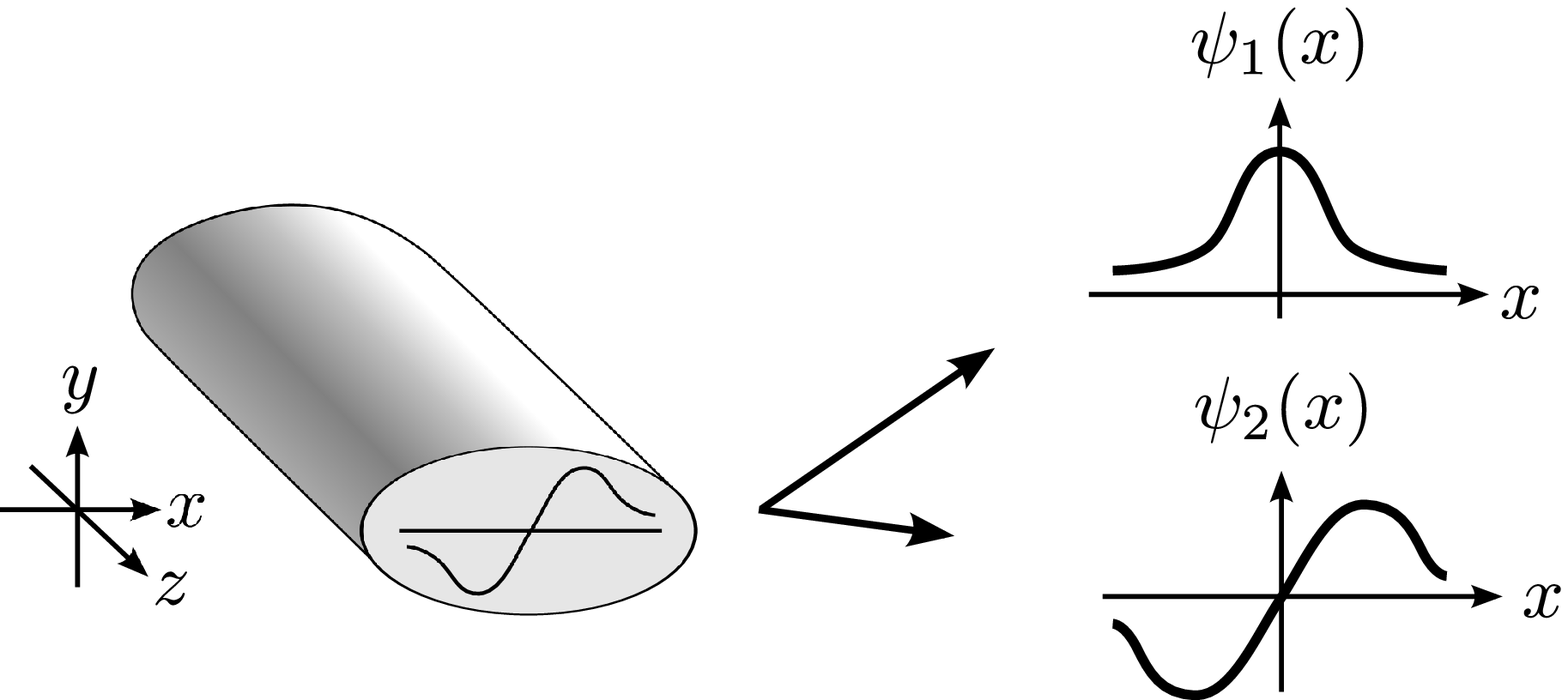}\\[0.75cm]
$(b)$\quad\includegraphics[scale=0.15]{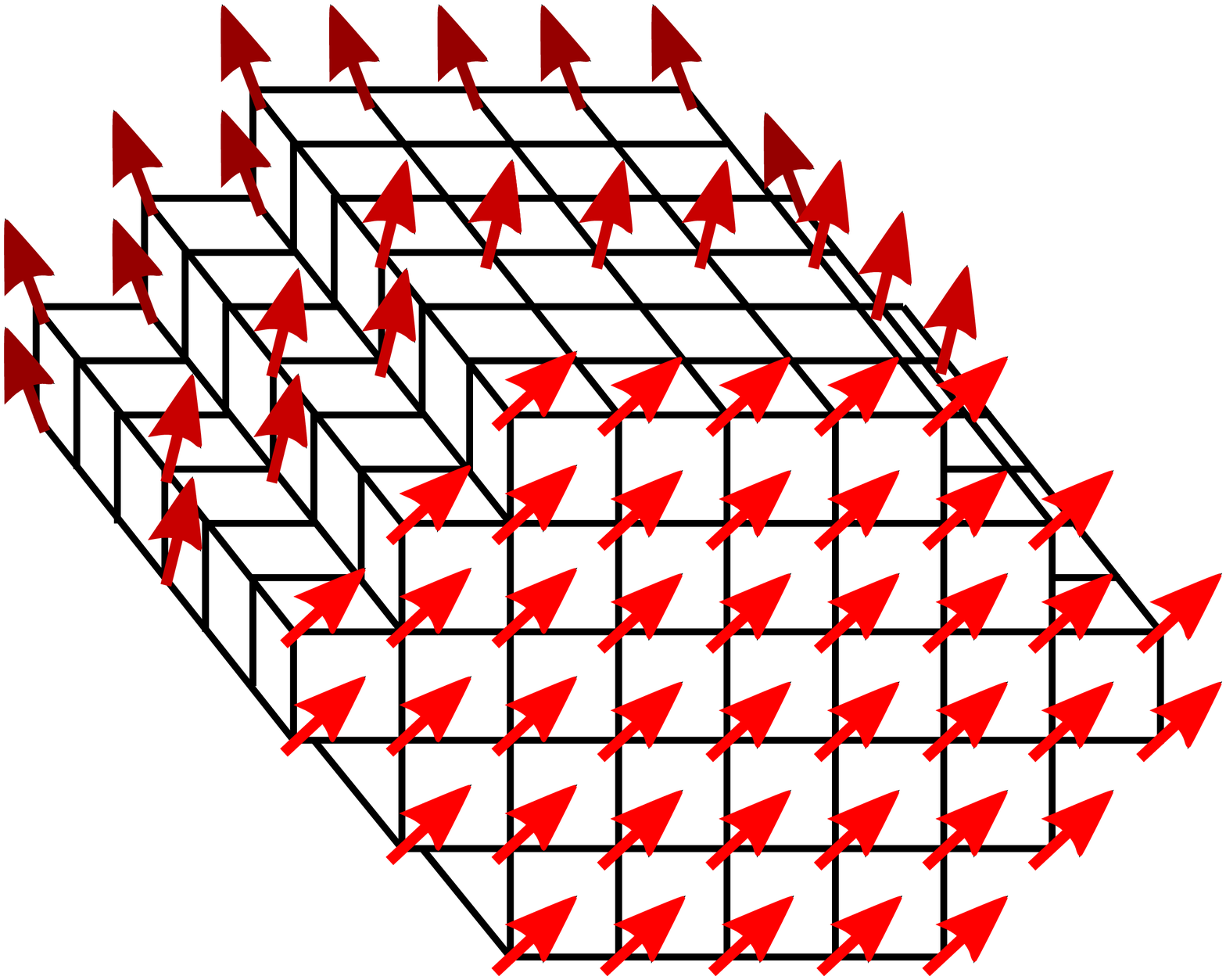}\quad\raisebox{1.45cm}{$\Rightarrow$}\quad\raisebox{0.5cm}{\includegraphics[scale=0.1]{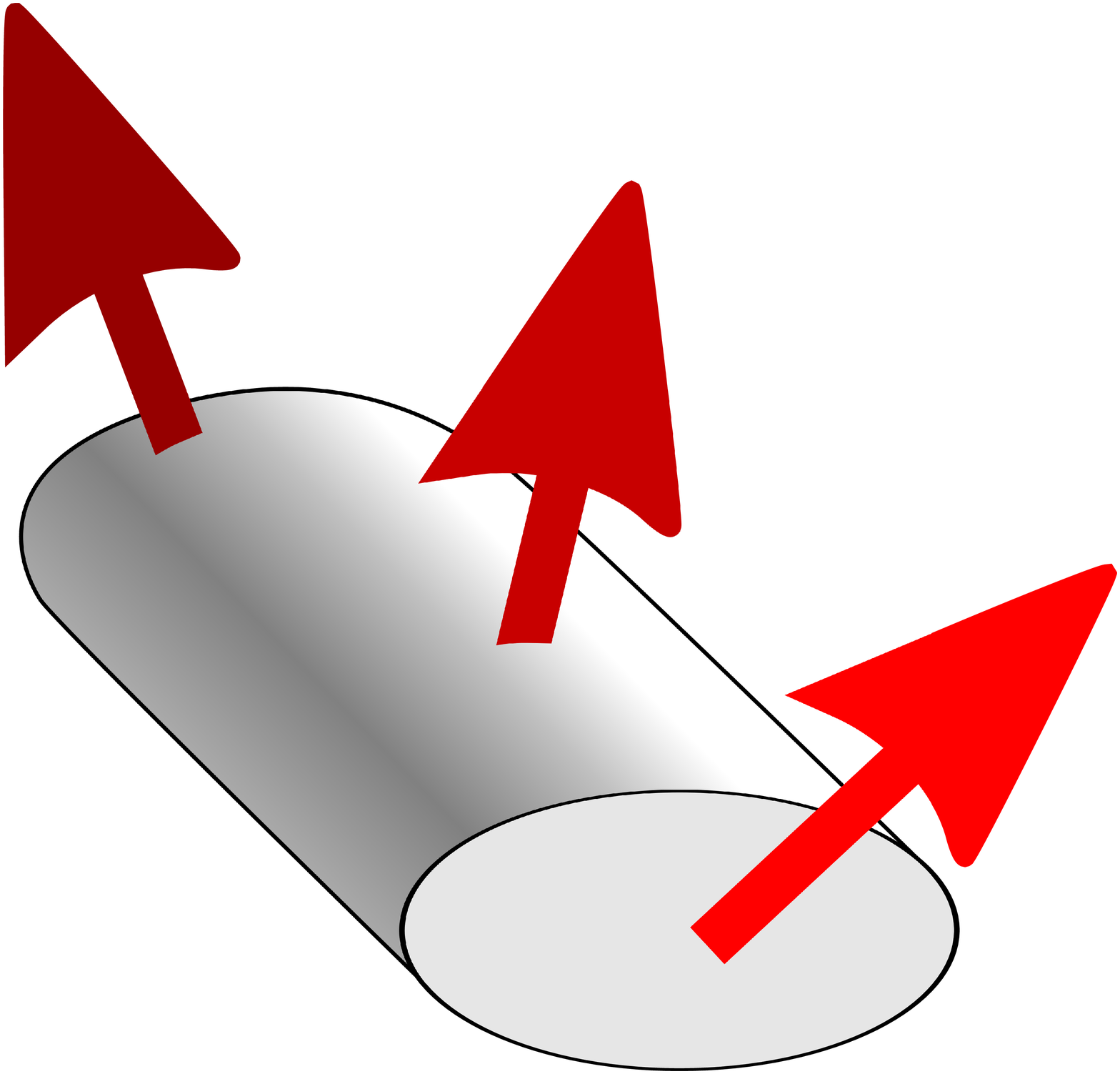}}
\caption{Panel $(a)$ depicts a quantum wire with two subbands defined in a two-dimensional electron gas by a harmonic confining potential. The subbands have transversal wave functions $\Psi_j(x,y) = \psi_j(x)\otimes\psi_1(y)$, and therefore exhibit different $x$-inversion symmetry with respect to the wire axis. A sketch of the nuclear spin lattice in the wire is given on the left-hand side of panel $(b)$ (for visibility reasons, not all spins are shown). Within a given cross-section, the nuclear spins align ferromagnetically, while they form helical orders along the wire direction (see Sec.~\ref{sec:feedback}). The right-hand side of panel $(b)$ shows the effective description of the nuclear spins as helical magnetic fields.}
\label{fig:wave_functions}
\end{figure}

The Hamiltonian is derived by first linearizing the spectrum around the Fermi points. Choosing the wire to be aligned along the $z$-axis, we decompose the annihilation operator for an electron of spin $\sigma = \uparrow,\downarrow$ in band $j=1,2$ as $c_{j\sigma}(z)=e^{i k_{Fj} z}R_{j\sigma}(z) + e^{-i k_{Fj} z}L_{j\sigma}(z)$, such that $e^{i k_{Fj}z}\,R_{j,\sigma}$ ($e^{-i k_{Fj}z}\,L_{j,\sigma}$) annihilates a right-moving (left-moving) particle ($k_{Fj}$ denotes the Fermi momentum in band $j$). This yields

\begin{subequations}\label{eq:total_hamilton_1}
\begin{align}
 H &= H_{\rm e} + H_{\rm en} ~,\\
 H_{\rm e} &=\sum_{j,\sigma} \, \int dz \,v_{Fj}\,\left(R_{j\sigma}^\dagger(-i\partial_z)R_{j\sigma}^\pdag-L_{j\sigma}^\dagger(-i\partial_z)L_{j\sigma}^\pdag\right)\nonumber\\
 &+\frac{U}{2}\,\int dz\, \rho_{\rm tot} \, \rho_{\rm tot} ~,\label{eq:ham_e}\\
 H_{\rm en} &= \sum_i A_0 \, \boldsymbol{S}_i\cdot \boldsymbol{I}_i ~,
\end{align}
\end{subequations}
where $v_{Fj}$ is the Fermi velocity of band $j$ and $U$ denotes the local interaction for the charge density $\rho_{\rm tot} = \sum_{j,\sigma} (R_{j\sigma}^\dagger R_{j\sigma}^\pdag + L_{j\sigma}^\dagger L_{j\sigma}^\pdag)$. The hyperfine coupling between the electron spin $\boldsymbol{S}_i$ and the nuclear spin $\boldsymbol{I}_i$ at site $i$ is $A_0$, where $i$ runs over all sites of the nuclear spin lattice within the support of the electronic wave functions. This includes typically $N_\perp \sim 50 \times 50$ sites in the transversal directions (the nuclear spin lattice is thus three-dimensional, see Fig.~\ref{fig:wave_functions}).\cite{pfeiffer_97} We finally assume that the wire has a finite length $L$, and that the lattice constant of the nuclear spins is $a$.

Since the Fermi energies $E_{Fj}$ in the two bands are typically much larger than the intrinsic energy scales of the nuclear spins as well as the coupling between nuclear spins and electrons, $E_{Fj}\gg A_0$,\cite{paget_hyperfine_gaas_77,braunecker_prb_09} the dynamics of electrons and nuclear spins decouple. The former mediate an effectively instantaneous RKKY interaction for the latter, while the nuclear spins act as a static magnetic Overhauser field for the electrons (in the disordered state, this Overhauser field vanishes). The separation of scales largely simplifies the following analysis.

\subsection{Effective one-dimensional model}\label{sec:1d_ham}
Because the electrons behave quasi one-dimensional, the coupling between nuclear spins and electrons is roughly the same for all spins within a given transversal section of the quantum wire. We can thus derive an effective quasi one-dimensional hyperfine Hamiltonian following Ref.~[\onlinecite{braunecker_prb_09}]. To this end, we note that the wave function of an electron can be written as a product of the longitudinal and the transversal state. For a particle at site $i = (i_{\parallel},i_{\perp})$, this implies $|\Psi_i\rangle = |i_{\parallel}\rangle \otimes |i_{\perp}\rangle$. The more appropriate basis for the electrons, however, is defined by the two subbands. Denoting the two transversal wave functions corresponding to the two occupied subbands $j=1,2$ as $\Psi_j(x,y)$, we find that the annihilation operator for an electron of spin $\sigma$ at site $i$ can be expressed as

\begin{subequations}
\begin{align}
 c_{i\sigma} &=  \sum_j \frac{\mathcal{C}_{ji_\perp}}{\sqrt{N_\perp}} \,c_{j \sigma i_\parallel} ~,\\
 \mathcal{C}_{ji_\perp} &= \sqrt{N_\perp}\,\int dx \, dy\,\Psi_j(x,y)\,\Phi_{i\perp}^*(x,y) ~,
\end{align}
\end{subequations}
where $\Phi_{i\perp}(x,y)$ is the Wannier orbital associated with $i_\perp$ (chosen such that $\mathcal{C}_{ji_\perp}$ is a real number). Within the support of the electronic wave functions, $\mathcal{C}_{ji_\perp}$ is of order one, while it vanishes outside the support. We can now express the hyperfine coupling in the two subband basis as

\begin{subequations}
\begin{align}
H_{\rm en} &= \frac{A_0}{N_\perp} \,\sum_{i_\parallel, j, j'}\,\boldsymbol{S}_{j j', i_\parallel} \cdot \boldsymbol{I}_{j j', i_\parallel} ~,\\
\boldsymbol{S}_{j j', i_\parallel} &= \sum_{\alpha \beta}\,c_{j\alpha i_{\parallel}}^\dagger\frac{\boldsymbol{\sigma}_{\alpha \beta}}{2}c_{j'\beta i_{\parallel}}^\pdag ~,\\
\boldsymbol{I}_{j j', i_\parallel} &= \sum_{i_\perp}  \mathcal{C}_{ji_\perp}\mathcal{C}_{j'i_\perp} \boldsymbol{I}_{i_\parallel i_\perp} ~,
\end{align}
\end{subequations}
where $\boldsymbol{\sigma}$ denotes the vector of Pauli matrices.

The terms $j=j'$ describe how the spin of electrons in band $j$ couples to the nuclear spins within a given section $i_\parallel$ of the wire. Since $(\mathcal{C}_{ji_\perp})^2 \sim 1$ across the section, the electron spin in fact rather couples to one big nuclear spin

\begin{align}
\boldsymbol{I}_{{\rm s}, i_\parallel} =\sum_{i_\perp}  (\mathcal{C}_{1i_\perp})^2 \boldsymbol{I}_{i_\parallel i_\perp} \approx \sum_{i_\perp}  (\mathcal{C}_{2i_\perp})^2 \boldsymbol{I}_{i_\parallel i_\perp}
\end{align}
than to $N_\perp$ individual nuclear spins. The interaction of an electron with a localized spin also allows for interband scattering processes. Since the two bands either correspond to the lowest two subbands of a single quantum wire or the symmetric and antisymmetric orbitals shared between two wires, the transversal wave functions $\Psi_1(x,y)$ and $\Psi_2(x,y)$ have different inversion symmetry with respect to the wire axis, see Fig.~\ref{fig:wave_functions}. We choose the first orbital to be the symmetric one,

\begin{equation}
 \Psi_1(x,y) = +\Psi_1(-x,y) ~~,~ \Psi_2(x,y) = -\Psi_2(-x,y) ~.
\end{equation}
The different inversion symmetry of the initial and final orbital of an interband spin scattering event implies that these processes couple to an inversion antisymmetric combination of the nuclear spins,

\begin{subequations}
\begin{align}
\boldsymbol{I}_{{\rm a}, i_\parallel} &=\sum_{i_\perp}  \mathcal{C}_{1i_\perp}\,\mathcal{C}_{2i_\perp}\, \boldsymbol{I}_{i_\parallel i_\perp} \\
&=\sum_{i_{\perp x},i_{\perp y}} |\,\mathcal{C}_{1i_\perp}\,\mathcal{C}_{2i_\perp}| \,\text{sgn}(i_{\perp x})\,\boldsymbol{I}_{i_\parallel i_{\perp x} i_{\perp y} }  ~.
\end{align}
\end{subequations}
With these definitions, the effective hyperfine Hamiltonian reads

\begin{align}\label{eq:hyperfine_ham}
H_{\rm en} &= \frac{A_0}{N_\perp} \,\sum_{i_\parallel}\,\left(\boldsymbol{S}_{11,  i_\parallel}+\boldsymbol{S}_{22,  i_\parallel}\right)\cdot \boldsymbol{I}_{{\rm s}, i_\parallel} \\
&+ \frac{A_0}{N_\perp} \,\sum_{i_\parallel}\, \left(\boldsymbol{S}_{12,  i_\parallel}+\boldsymbol{S}_{21,  i_\parallel}\right)\cdot \boldsymbol{I}_{{\rm a}, i_\parallel} ~.\nonumber
\end{align}
The electrons thus couple to either a symmetric or antisymmetric combination of all nuclear spins within a given section of the wire, depending on whether the spin scattering event changes the transversal wave function or not. These effective spins have the size $|\boldsymbol{I}_{{\rm s},i_\parallel}|, |\boldsymbol{I}_{{\rm a},i_\parallel}| \sim I N_\perp$, where $I$ is the length of the individual nuclear spins. Because $N_\perp$ is typically of the order of (a few) thousand, these effective nuclear spins can be treated semiclassically, such that for instance Kondo correlations can be neglected. The reduced coupling strength $A_0/N_\perp$ finally reflects the spread of the electronic wave function across the wire section. 

\section{Helical order of the nuclear spins: possible scenarios}\label{sec:helix_order}
Like for the single subband quantum wire,\cite{braunecker_prb_09} we use the separation of scales between the nuclear spins and the electrons to step by step derive a self-consistent description of the coupling between electrons and nuclear spins. We start from two gapless Luttinger liquids and a disordered bath of slow nuclear spins. By virtue of the hyperfine coupling, the electrons mediate an RKKY interaction for both nuclear spin superpositions $\boldsymbol{I}_{\rm s}$ and $\boldsymbol{I}_{\rm a}$. As detailed in Appendix \ref{append:rkky}, the RKKY interaction is described by the Hamiltonian

\begin{figure}
\centering
\includegraphics[scale=0.5]{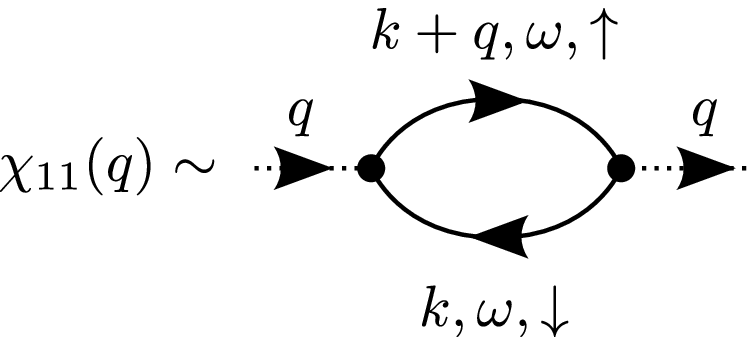}\quad\quad\quad\includegraphics[scale=0.5]{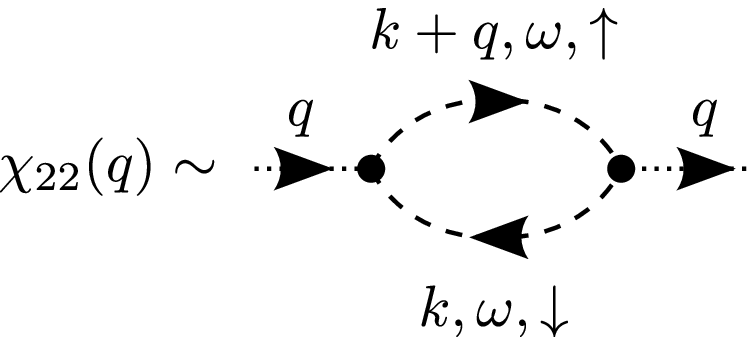}\\[0.2cm]
\includegraphics[scale=0.5]{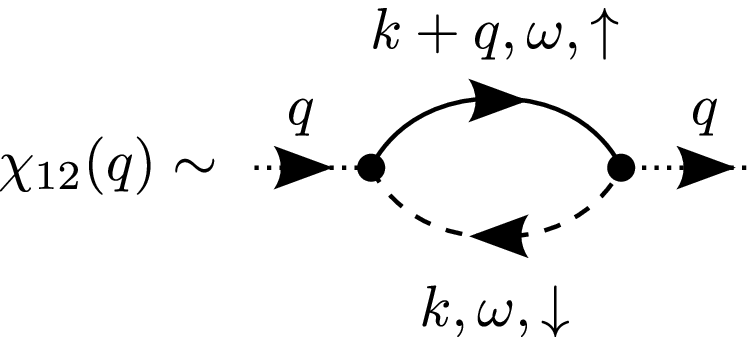}
\caption{Fermionic diagrams contributing to the ($x$ and $y$ components of the) spin susceptibilities $\chi_{ij}$. Solid lines denote a particle in band 1 at, e.g., momentum $k+q$, frequency $\omega$ and spin up, dashed lines denote a particle in band 2 (note that the Luttinger liquid approach yields the density-density interacting versions of these diagrams). The dotted lines denote the momentum $q$ emitted/absorbed by the nuclear spins.}
\label{fig:susceptibility_bubbles}
\end{figure}
\begin{figure}
\centering
\includegraphics[scale=0.4]{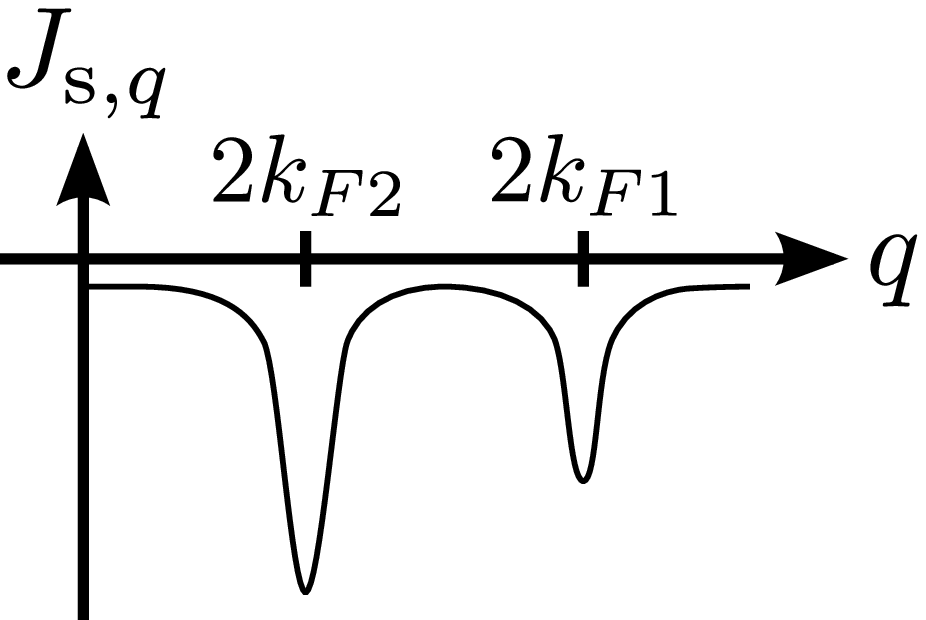}\quad\quad\includegraphics[scale=0.4]{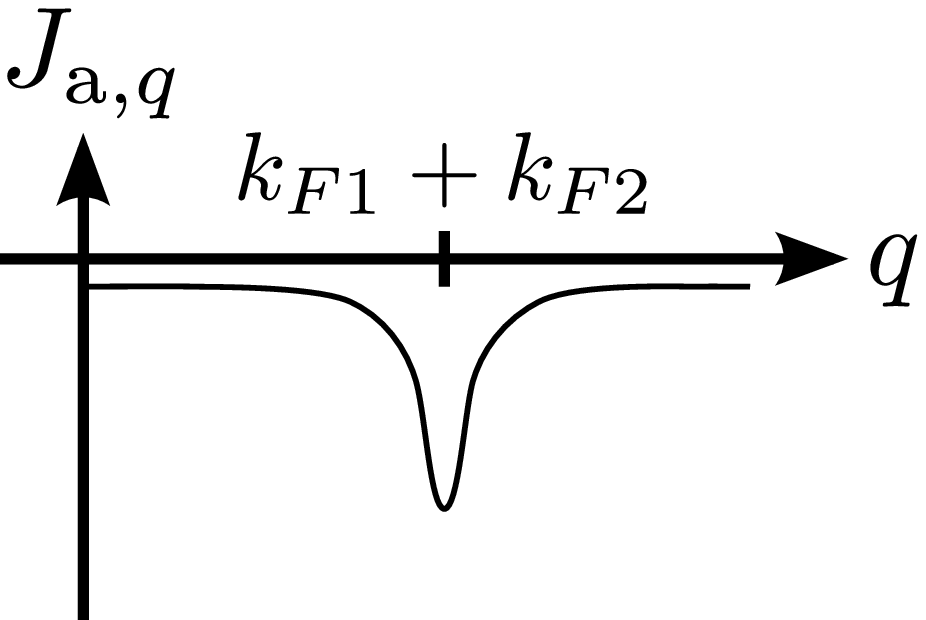}
\caption{RKKY exchange interactions $J_{\rm s}$ and $J_{\rm a}$ for $g_{22} < g_{12} < g_{11}$ and at finite temperature.}
\label{fig:ja_js}
\end{figure}


\begin{align}
H_{\rm RKKY} &= \frac{1}{N} \sum_{q,\alpha,\beta}\Bigl(I_{{\rm s},-q}^\alpha \,\frac{J_{{\rm s},q}^{\alpha\beta}}{N_\perp^2}\, I_{{\rm s},q}^\beta+ I_{{\rm a},-q}^\alpha \,\frac{J_{{\rm s},a}^{\alpha\beta}}{N_\perp^2}\,\, I_{{\rm a},q}^\beta\Bigr)~,\label{eq:H_rkky}
\end{align}
where the exchange interactions are determined by the static part of the respective spin susceptibilities, $J_{(\cdot),q}^{\alpha\beta} = A_0^2/(2N)\,\chi_{(\cdot),q}^{\alpha\beta,\rm ret}(\omega\to0)$, and where $N=L/a$ is the number of longitudinal lattice sites (more precisely, $\chi_{(\cdot),q}^{\alpha\beta,\rm ret}$ is the $(\alpha,\beta)$ component of the retarded spin susceptibility tensor in the $(\cdot)=\mathrm{s,a}$ channel, $q$ denotes the one-dimensional momentum along the wire axis and $\omega$ is the frequency). Like in two-dimensional electron gases,\cite{2deg_spin_suscept_zak} the susceptibilities correspond to the interacting versions of the diagrams shown in Fig.~\ref{fig:susceptibility_bubbles} (the Luttinger liquid formalism however spares us lengthy resummation schemes). Similar to the single subband case,\cite{braunecker_prb_09} and as we will detail below, particle-hole bubbles within each band and between the bands result at zero temperature in divergences of the RKKY exchange couplings at momenta $2 k_{F1}$ and $2k_{F2}$ in the symmetric channel, and at $k_{F1}+k_{F2}$ in the antisymmetric channel (provided that the exponents defined in the following Eq.~\eqref{eq:rkky_scaling_exponents} fulfill $2g_{ij} < 2$),

\begin{subequations}\label{eq:rkky_scaling_exponents}
\begin{align}
 &\left.J_{{\rm s},q}^{\alpha\beta}\right|_{q\approx \pm 2k_{F1}} \sim \delta_{\alpha\beta}\,\left|q\mp 2 k_{F1}\right|^{2g_{11}-2}~,\\
 &\left.J_{{\rm s},q}^{\alpha\beta}\right|_{q\approx \pm 2k_{F2}} \sim \delta_{\alpha\beta}\,\left|q\mp 2 k_{F1}\right|^{2g_{22}-2}~,\\
 &\left.J_{{\rm a},q}^{\alpha\beta}\right|_{q\approx \pm (k_{F1}+k_{F2})} \sim \delta_{\alpha\beta}\,\left|q\mp (k_{F1}+k_{F2})\right|^{2g_{12}-2}~.
\end{align}
\end{subequations}
We note that the exchange couplings are diagonal in spin space because the Hamiltonian preserves the total spin. At finite temperatures, the divergences turn into sharp dips, whose depth is controlled by the temperature $T$ and the exponents $g_{ij}$, and whose width is of the order of the thermal momentum, see Fig.~\ref{fig:ja_js}.

In order to gain energy, the slow nuclear spins will now orient into the minima of the susceptibilities. As discussed in Refs.~[\onlinecite{braunecker_prl_09}] and [\onlinecite{braunecker_prb_09}], this results in helically ordered states at the momenta corresponding to the dips of the susceptibilities. In a two-subband quantum wire, the presence of three different combinations of Fermi momenta $k_{Fi}+k_{Fj}$ could a priori result in a superposition of the three helices,

\begin{subequations}
\begin{align}
 \boldsymbol{I}_{{\rm s},i_{\parallel}} &=  I N_\perp \left[m_{2k_{F1}}\begin{pmatrix}\cos(2 k_{F1} z_i)\\\pm\sin(2 k_{F1} z_i)\\0\end{pmatrix}\right. \\
 &+ \left.m_{2k_{F2}} \begin{pmatrix}\cos(2 k_{F2} z_i)\\\pm\sin(2 k_{F2} z_i)\\0\end{pmatrix}\right]\nonumber ~,\\
 \boldsymbol{I}_{{\rm a},i_{\parallel}} &=I N_\perp\, m_{k_{F1} + k_{F2}}\begin{pmatrix}\cos((k_{F1} + k_{F2}) z_i)\\\pm\sin((k_{F1} + k_{F2}) z_i)\\0\end{pmatrix} ~,
\end{align}
\end{subequations}
where we used $z_i = a \, i_\parallel$. In the semiclassical approximation employed here, the three magnetizations correspond to the fraction of microscopic nuclear spins participating in either of the three orders. At zero temperature, when thermal fluctuations do not weaken the order, the magnetizations have to add up to one, $m_{2k_{F1}} + m_{2k_{F2}} +  m_{k_{F1} + k_{F2}} = 1$. The helicities and ordering planes are in principle spontaneously and independently chosen for each of the three helices. This analysis however neglects the feedback between nuclear spins and electrons and is thus not self-consistent. Any finite nuclear spin polarization acts back onto the electrons as a static magnetic Overhauser field. Since these fields form at the momenta $2k_{F1}$, $2k_{F2}$ or $k_{F1}+k_{F2}$, they allow for backscattering within one band or between the two bands. Due to the different inversion symmetries of the electronic wave functions in the two bands, however, intraband (interband) backscattering is only possible in the symmetric (antisymmetric) channel. The backscattering in turn opens up gaps in the electronic spectrum (see Sec.~\ref{subsec:low_t}). If however the electrons are gapped, they cannot mediate the RKKY interaction needed to establish the nuclear spin order in the first place. 

\begin{figure}
\centering
$(a)$\quad\includegraphics[scale=0.35]{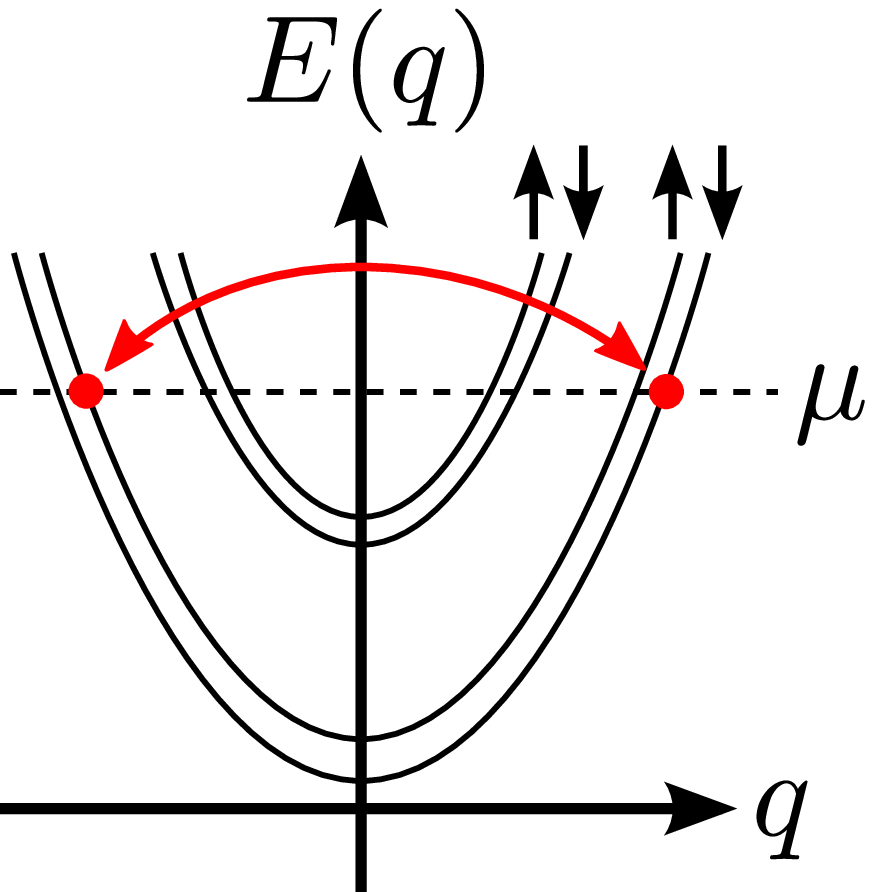}\quad\raisebox{1.25cm}{$\Rightarrow$}\quad\includegraphics[scale=0.35]{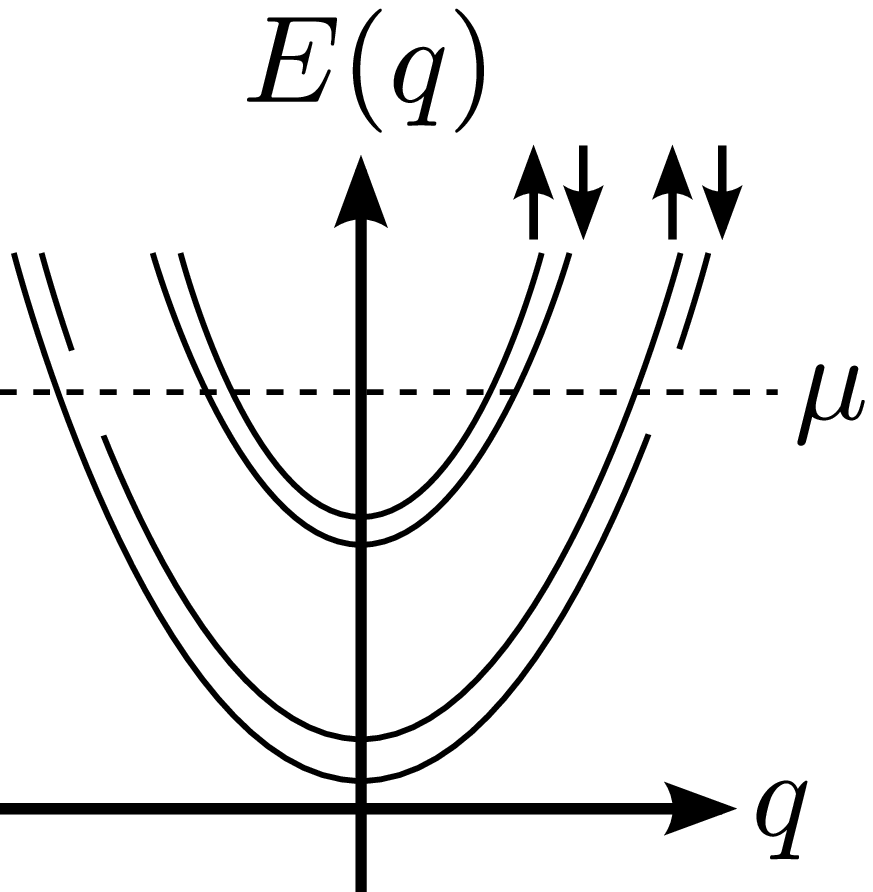}\\
$(b)$\quad\includegraphics[scale=0.35]{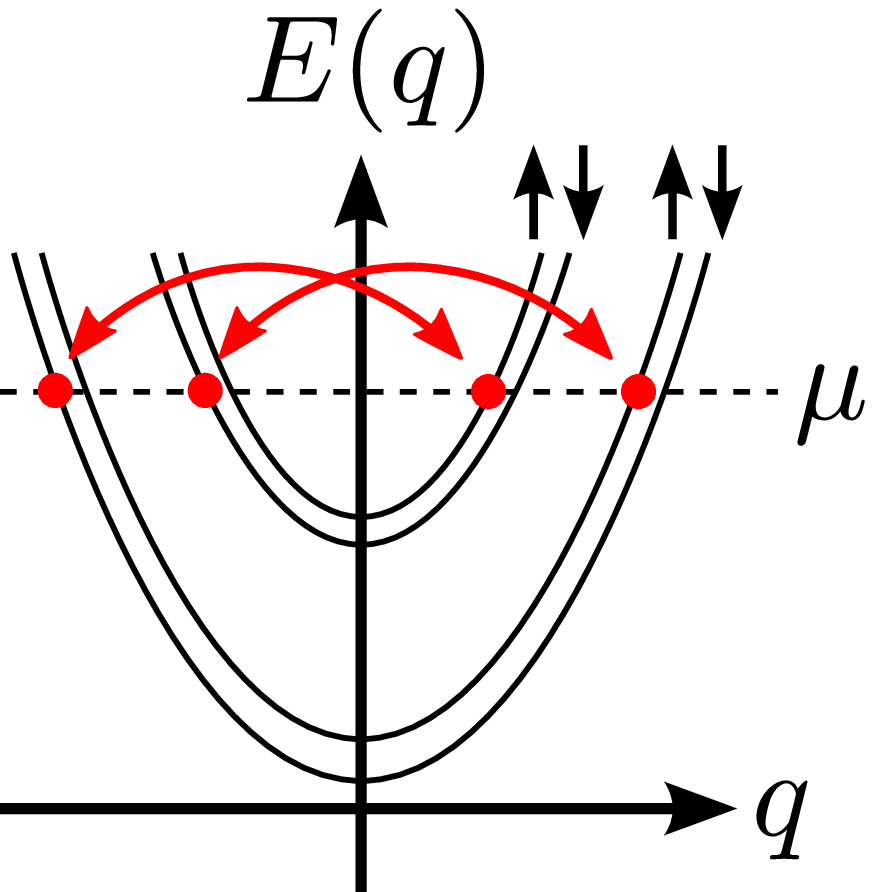}\quad\raisebox{1.25cm}{$\Rightarrow$}\quad\includegraphics[scale=0.35]{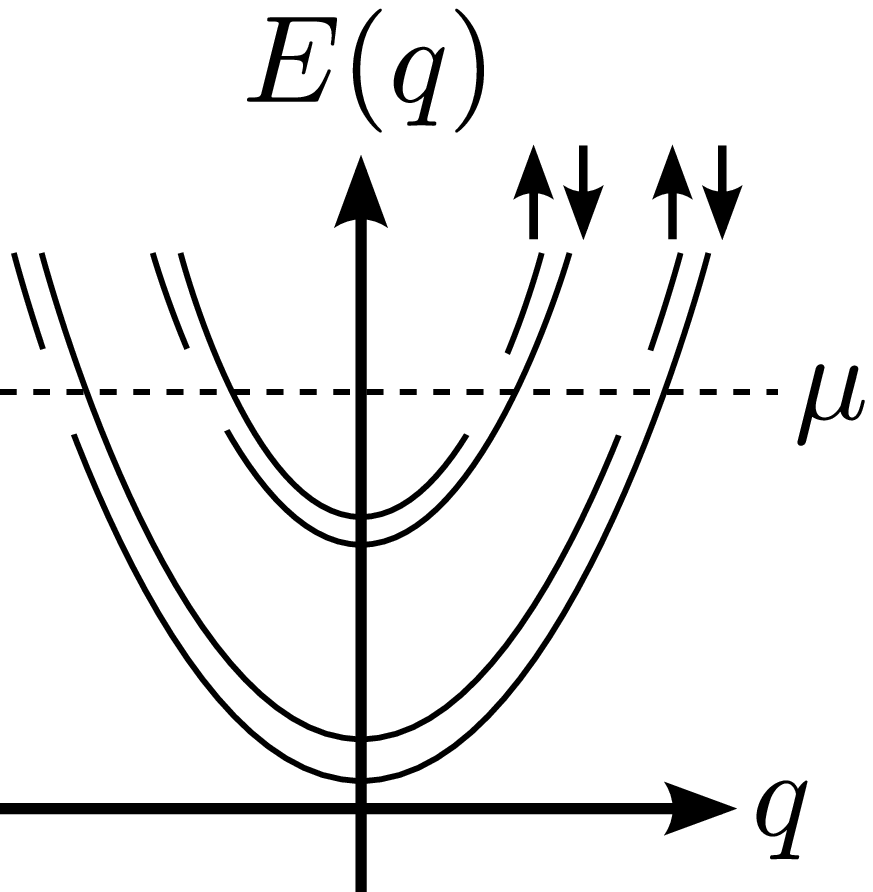}
\caption{Opening of gaps around the chemical potential $\mu$ in the dispersion $E(q)$ due to the magnetic Overhauser field in a two-subband quantum wire. Panel $(a)$ depicts the effect of a helical field of positive helicity and momentum $2k_{F1}$ in the symmetric channel, which allows for scattering between right moving spin down particles and left moving spin up particles. The scattering gaps out these two modes. Panel $(b)$ shows how a helical field in the antisymmetric channel of negative helicity and momentum $k_{F1}+k_{F2}$ gaps out right moving spin up particles and left moving spin down particles in both bands. As explained in the main text, the intraband order depicted in panel (a) and the interband order shown in panel (b) are mutually exclusive because their combination would gap out the entire lower subband.}
\label{fig:scattering_gaps}
\end{figure}

To find the possible self-consistent orders of nuclear spins and electrons, it is useful to note that a helix in the symmetric channel at $2k_{F1}$ ($2k_{F2}$), for concreteness of positive helicity, allows only for backscattering between right moving spin down and left moving spin up particles within the first (second) band. Such a nuclear spin polarization will thus gap out the right moving spin down and left moving spin up modes in the first (second) band, see Fig.~\ref{fig:scattering_gaps}$(a)$ and Sec.~\ref{subsec:low_t}. The left moving spin down and right moving spin up, on the other hand, remain unaffected. Similarly, we find that an Overhauser field in the antisymmetric channel at $k_{F1} + k_{F2}$ (for concreteness of negative helicity) allows for backscattering of right moving spin up in both bands and left moving spin down in both bands. It can therefore open up gaps in the two bands, as depicted in Fig.~\ref{fig:scattering_gaps}$(b)$. These gaps correspond to the formation of an electron spin helix in phase with the nuclear spin helix (or out of phase for antiferromagnetic hyperfine coupling). The Overhauser field induced gaps in the electronic spectrum are relevant in the renormalization group (RG) sense, again provided that the associated exponent fulfills $2g_{ij} < 2$ (see Sec.~\ref{subsec:low_t}). Since $g_{ij}$ is intimately related to the Luttinger liquid parameters, interactions are crucial for this strong renormalization of the Overhauser gaps.

In the case of a single helix (be it at $2 k_{F1}$, $2k_{F2}$ or $k_{F1} +k_{F2}$), the remaining ungapped electronic modes provide the residual RKKY interaction needed to self-consistently establish the nuclear spin order. For a superposition of helices, one finds by counting of the gapless right and left moving modes that a self-consistent helical order can only occur if either helices at $2k_{F1}$ and $2k_{F2}$ coexist (or if there is just one of them) while there is no helix in the antisymmetric channel, or  if there is a helix in this latter channel at momentum $k_{F1}+k_{F2}$ and no helices in the symmetric channel. As an example, consider the effectively forbidden coexistence of a helix of positive helicity in the symmetric channel at $2k_{F1}$ and a helix of negative helicity in the antisymmetric channel at $k_{F1}+k_{F2}$ (see Fig.~\ref{fig:scattering_gaps}). This combination of Overhauser fields would gap out the entire spectrum in the first band, plus parts of the spectrum in the second band. Since the first band is fully gapped, scattering events involving this band are strongly suppressed. This would eliminate the minima at $\pm 2k_{F1}$ and at $\pm (k_{F1}+k_{F2})$ in the RKKY interactions and thus render the analysis inconsistent. In addition, the nuclear spins would then reorient into the remaining minima at $\pm 2k_{F2}$, leading to a strong reduction of the magnetic Overhauser fields at $2k_{F1}$ (symmetric channel) and $k_{F1}+k_{F2}$ (antisymmetric channel). This would finally suppress the electronic gaps due to these fields, which is energetically strongly unfavorable. The coexistence of a helix of positive helicity in the symmetric channel at $2k_{F1}$ and a helix of negative helicity in the antisymmetric channel at $k_{F1}+k_{F2}$ can therefore be excluded. Similar reasonings can be made for other combinations of Overhauser fields. Coexistence of helices of equal helicity in the symmetric and antisymmetric channel, on the other hand, would lead to competing ordering mechanism for the electrons. We thus find that the orders of the nuclear spins in the symmetric channel $\langle \boldsymbol{I}_{\rm s}\rangle$ and the antisymmetric channel $\langle \boldsymbol{I}_{\rm a}\rangle$ are mutually exclusive if the feedback between electrons and nuclear spins is taken into account.

\section{Self-consistent solution with feedback effects}
\label{sec:feedback}
The ground state of the quantum wire depends on the interplay of electrons and nuclear spins, and has to be determined as the minimum of the total energy. It should thus take into account the energy gain of the nuclear spins due to the ordering into the (self-consistently present) minima of the RKKY interaction, as well as the energy gain of the electrons due to the opening of gaps.  

\subsection{Dominant ordering mechanism: high temperature analysis}\label{subsec:high_t}
To identify the dominant ordering mechanism of the nuclear spins (symmetric or antisymmetric channel), we first analyze the onset of nuclear spin order upon lowering the temperature. At sufficiently high temperatures, the nuclear spins are certainly disordered. Using $|\boldsymbol{I}_{\rm s, a}|/N_{\perp} \sim I$, we infer from the RKKY Hamiltonian given in Eq.~\eqref{eq:H_rkky} that the typical temperature scale for the onset of nuclear spin order in the (anti-) symmetric channel at momentum $q$ is $J_{\rm s(a),q}\,I^2$. The critical temperature will be discussed in more detail in Sec.~\ref{sec:order_and_finite_t}. Just before the onset of the nuclear spin order, when there is no Overhauser field present yet, the RKKY interaction mediated by the fully ungapped electron sector can be derived from the electron Hamiltonian given in Eq.~\eqref{eq:ham_e}. In order to include the density-density interactions, we treat this Hamiltonian with standard bosonization techniques, which allow for its full diagonalization, see Appendix \ref{append:diag}. This diagonal form finally allows the determination of the spin susceptibilities and thus the RKKY exchange couplings, as detailed in Appendix \ref{Appendix:suscept}. 

\begin{figure}
\centering
\includegraphics[scale=1.0]{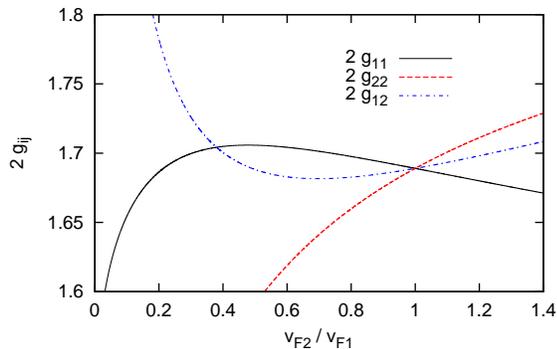}
\caption{Exponents $2 g_{ij}$ of the RKKY exchange couplings, see Eq.~\eqref{eq:rkky_scaling_exponents}, as a function of $v_{F2}/v_{F1}$ for fixed $v_{F1}$ and interaction strength $U$. These exponents describe the regime of disordered nuclear spins, e.g.~at sufficiently high temperatures. Consequently, they do not contain any feedback effect between electrons and nuclear spins, see Sec.~\ref{sec:feedback}.}
\label{fig:scaling_exponents_small}
\end{figure} 

\begin{figure}
\centering
\includegraphics[scale=1.0]{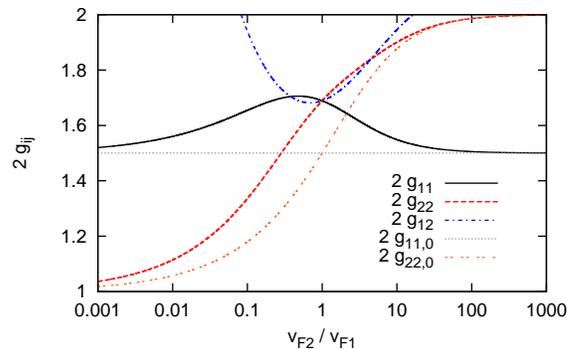}
\caption{Exponents $2 g_{ij}$ of the RKKY exchange couplings, see Eq.~\eqref{eq:rkky_scaling_exponents}, as a function of $v_{F2}/v_{F1}$ for fixed $v_{F1}$ and interaction strength $U$.  On this larger scale logarithmic plot, we also show the exponents $2g_{ii,0}$ for decoupled bands, see main text. Like for the smaller scale plot in Fig.~\ref{fig:scaling_exponents_small}, the nuclear spins are assumed to be disordered.}
\label{fig:scaling_exponents_large}
\end{figure}

The resulting exponents $2g_{ij}$ defined in Eq.~\eqref{eq:rkky_scaling_exponents}, which control the strength of the RKKY interaction, are plotted in Figs.~\ref{fig:scaling_exponents_small} and \ref{fig:scaling_exponents_large} as a function of $v_{F2}/v_{F1}$ for fixed $v_{F1}$ and fixed interaction strength $U$ (thus following the filling of the second subband). The interaction strength is chosen such that the Luttinger liquid parameter in the charge sector of the first subband is $K_{c1} = 0.5$, while we use $K_{si} = 1$ because the interactions in the spin sectors are assumed to be much smaller than the ones in the charge sectors (in agreement with experiments).\cite{steinberg_charge_fractionalization_08, jompol_kc_ks} We find that at the special point $v_{F1} = v_{F2}$, all susceptibilities have the same exponent. This has already been discussed in the context of Carbon nanotubes,\cite{braunecker_prb_09} where the two subbands correspond to the two inequivalent Dirac cones. Away from this special point,
 however, the exchange coupling $J_{\rm s}$ associated with the symmetric superposition of nuclear spins $\boldsymbol{I}_{\rm s}$ is always more singular than the exchange coupling $J_{\rm a}$ associated with $\boldsymbol{I}_{\rm a}$ (we recall that $J_{\rm s} \sim \chi_{11} + \chi_{22}$ while $J_{\rm a} \sim \chi_{12}+\chi_{21}$). Therefore, the system will initially order in the symmetric channel.

Since the spin susceptibilities and thus the RKKY exchange couplings are mediated by particle-hole pairs exchanged between two nuclear spins, see Fig.~\ref{fig:susceptibility_bubbles}, it is physically not surprising that the interband RKKY exchange can only be effective if the particle and the hole propagate at comparable velocities. Consequently, the two subbands are essentially decoupled when $v_{F1} \gg v_{F2}$ or $v_{F1} \ll v_{F2}$. This is shown in the larger scale plot in Fig.~\ref{fig:scaling_exponents_large}, which indicates that the exponents $2g_{ii}$ approach their decoupled values $g_{11,0}$ and $g_{22,0}$ (calculated in the absence of interband density interactions) away from $v_{F1} \approx v_{F2}$. Only close to $v_{F1}=v_{F2}$, the interband channel participates in the RKKY interaction. At the same time, the weight of the intraband exchange channels decreases and the corresponding exponents $g_{11}$ and $g_{22}$ are larger than in the decoupled case. A related situation arises in Coulomb drag setup, where the drag is most efficient if the Fermi velocities of the two wires are approximately equal.\cite{drag_literature}

\subsection{Self-consistent low temperature ground state}\label{subsec:low_t}
Given that the dominant instability at high temperatures occurs in the symmetric sector, we calculate the low temperature ground state \textit{assuming} that the nuclear spins form helices in the symmetric channel. This assumption will be justified a posteriori in Sec.~\ref{sec:inter_intra}. We start from a general superposition of nuclear spin helices in the symmetric channel,

\begin{align}\label{eq:actual_overhauser_field}
 \langle\boldsymbol{I}_{{\rm s},i_{\parallel}}\rangle &=  I N_\perp \left[m_{2k_{F1}}\begin{pmatrix}\cos(2 k_{F1} z_i)\\\sin(2 k_{F1} z_i)\\0\end{pmatrix}\right. \\
 &+ \left.m_{2k_{F2}} \begin{pmatrix}\cos(2 k_{F2} z_i)\\\sin(2 k_{F2} z_i)\\0\end{pmatrix}\right]\nonumber ~,
\end{align}
with magnetizations $m_{2k_{F1}}+m_{2k_{F2}} = 1$ at zero temperature, and subsequently minimize the total energy with respect to the magnetizations $m_{2k_{F1}}$ and $ m_{2k_{F2}}$. The system is analyzed in a bosonized language, in which the annihilation operator for a particle of spin $\sigma$ in band $i$ reads

\begin{align}
r_{i\sigma}(z) = \frac{U_{ri\sigma}}{\sqrt{2\pi\alpha}}\,e^{-i(r\phi_{i\sigma}(z)-\theta_{i\sigma}(z))} ~,
\end{align}
where $r=R,L\equiv+,-$, the corresponding Klein factors are denoted as $U_{ri\sigma}$, and $\alpha$ is a short distance cutoff (here taken to be the lattice spacing). The bosonic fields $\phi_{i\sigma}$ and $\theta_{i\sigma}$ fulfill the standard commutation relation $[\phi_{i\sigma}(z),\theta_{i'\sigma'}(z')]=\delta_{ii'}\delta_{\sigma\sigma'}\,(i\pi/2)\ \text{sgn}(z'-z)$.\cite{giamarchi_book} It is furthermore convenient to introduce spin and charge degrees of freedom via the canonical transformation $\phi_{i\hspace*{-2.5pt}\begin{array}{c}\\[-16.5pt] {}_{c}\\[-7pt] {}_{s}\end{array}}(z) = (\phi_{i\uparrow}\pm\phi_{i\downarrow})/\sqrt{2}$ and $\theta_{i\hspace*{-2.5pt}\begin{array}{c}\\[-16.5pt] {}_{c}\\[-7pt] {}_{s}\end{array}}(z) = (\theta_{i\uparrow}\pm\theta_{i\downarrow})/\sqrt{2}$.

\begin{figure}
\centering
$(a)$\quad\includegraphics[scale=0.75]{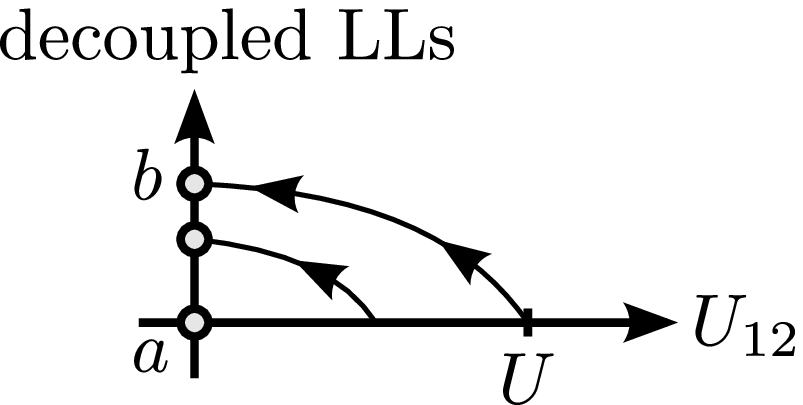}\\[0.2cm]
$(b)$\quad\includegraphics[scale=0.75]{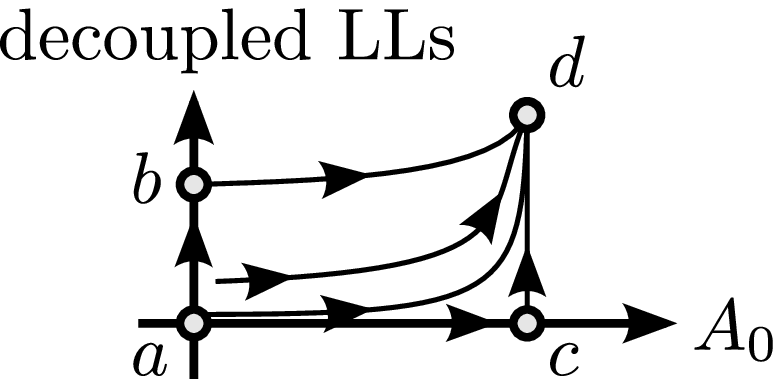}
\caption{RG flow of the two-subband quantum wire system. The upper panel $(a)$ shows the flow for density-density coupled subbands in the absence of hyperfine coupling. For any value of the interband interaction $U_{12}$, the system can be described by 2 decoupled spinful Luttinger liquids (LL), characterized however by different Luttinger liquid parameters and velocities. The lower panel $(b)$ shows how the flow is modified due to the presence of the hyperfine coupling and a resulting ordering of the nuclear spins in the symmetric channel. Here, the interband interaction would correspond to the out-of-plane axis (not shown), and panel $(b)$ is the projection of the full flow into the given plane.}
\label{fig:rg_fixed_points}
\end{figure}

As shown in Appendix \ref{append:overhauser_fields}, the hyperfine coupling between the ordered nuclear spins and the electrons yields various cosine-terms. If the Fermi momenta $k_{F1}$ and $k_{F2}$ are non-commensurate, the only non-oscillating terms read

 \begin{align}
H_{\rm en}^{\rm non-osc.} =\sum_{i=1,2} \frac{B_{xy,i}}{2 \pi \alpha} \int dz\,\cos\left(\sqrt{2}(\phi_{ic} + \theta_{is})\right) \text{ ,}
\end{align}
where $B_{xy,i} = I \, A_0\, m_{2k_{Fi}}$. These Overhauser fields are RG relevant and thus lead to an ordering of $\phi_{ic} + \theta_{is}$, i.e.~half of the electronic degrees of freedom. This in turn feeds back to the nuclear spins by a modification of the RKKY exchange interaction, which is now only mediated by the gapless part of the electronic spectrum. To include this feedback effect into the theory, we start from the non-diagonal version of the electron Hamiltonian in the presence of the Overhauser fields and interband interactions,

\begin{align}\label{eq:el_ham_ov}
H_{\rm e} &= \sum_{i=1,2}\sum_{j=c,s}\int \frac{dz}{2\pi}\,\left(\frac{u_{ij}}{K_{ij}}\,(\partial_z \phi_{ij})^2 + u_{ij} K_{ij} \, (\partial_z\theta_{ij})^2\right)\nonumber\\
 &+ \int \frac{dz}{2\pi}\,\frac{4U}{\pi}\, (\partial_z \phi_{1c})(\partial_z \phi_{2c})\\
  &+\sum_{i=1,2} \frac{B_{xy,i}}{2 \pi \alpha} \int dz\,\cos\left(\sqrt{2}(\phi_{ic} + \theta_{is})\right)~.\nonumber
\end{align}
Here, $u_{ij}$ and $K_{ij}$ are the effective velocities and Luttinger liquid parameters in the spin and charge sectors of the two bands with $u_{ij} = v_{Fi}/K_{ij}$. This representation indicates that the interband Coulomb interaction and the intraband Overhauser fields have competing effects and should be treated simultaneously. While the interband interaction is diagonalized by band-mixed boson fields given in Eq.~\eqref{eq:diagonalization_trafo}, the magnetic Overhauser fields induce an intraband ordering and gap out parts of the spectrum within each band, which opposes a complete mixing of the bosonic fields. Fig.~\ref{fig:rg_fixed_points} depicts this in terms of RG fixed points. The first line of Eq.~\eqref{eq:el_ham_ov} corresponds to the fixed point $a$ with two decoupled Luttinger liquids. Adding an interband interaction $U_{12} = U$ but no coupling to the nuclear spins takes the system to a different fixed point $b$ described by 2 decoupled Luttinger liquids with modified Luttinger parameters and velocities whose Hamiltonian is given in Eq.~\eqref{eq:diag_el_ham}. This flow is shown in the upper panel of Fig.~\ref{fig:rg_fixed_points}.  The fixed point $b$ has been used to calculate the spin susceptibilities in Sec.~\ref{subsec:high_t}. The lower panel of Fig.~\ref{fig:rg_fixed_points}, on the other hand, depicts how the RG flow is changed due to the hyperfine coupling $A_0$ to the nuclear spins. Through the Overhauser fields, the hyperfine coupling gaps out half of the electronic degrees of freedom. As a consequence, the fixed points $a$ and $b$ are not stable at low temperatures. If there were no interband interaction, the system would be described by a fixed point $c$ with 2 gapless and 2 gapped Luttinger liquids. Now adding an interband coupling finally takes the system to yet another fixed point $d$, which is the one we are after.

Given that the intraband Overhauser fields are strongly RG relevant, they dominate the initial RG flow compared to the marginal interband density-density interaction. The latter is in any case only important when $v_{F1}\approx v_{F2}$ and may then indeed slightly modify the scaling of the Overhauser gap, but not the presence of an ordered state.  We thus first describe the flow from fixed point $a$ to fixed point $c$ and subsequently take into account the interband interaction. Close the to fixed point $a$, the term $\cos(\sqrt{2}(\phi_{ic} + \theta_{is}))$ has the scaling dimension $g_i = (K_{ic}+1/K_{is})/2$. Parameterizing the running short distance cutoff as $\alpha(b) = b\,\alpha$, the flow of $B_{xy,i}$ between the fixed points $a$ and $c$ reads

\begin{align}
\frac{\partial B_{xy,i}}{\partial\log(b)} = (1-g_{i})\,B_{xy,i}~.
\end{align}
As discussed in Appendix \ref{append:first_flow} (and similar to the single subband case),\cite{braunecker_prb_09} this leads to gaps $\Delta_{i+}$ for the bosonic fields $\phi_{i+} \sim \phi_{ic} + \theta_{is}$. There are however two more bosonic modes $\phi_{i-}$ resulting from the canonical transformation $\phi_c, \theta_s \to \phi_+, \phi_-$. These modes do not couple to the Overhauser fields and remain gapless. 

Expanding the cosine-terms to second order\cite{giamarchi_book} and using $u_{i+} = (u_{ic} K_{ic}+u_{is}/K_{is})/(K_{ic}+1/K_{is})$ and $K_{i} = 2 g_{i} = K_{ic}+1/K_{is}$ yields

\begin{align}
\Delta_{i+}(b) = \sqrt{\frac{K_{i}\,u_{i+}\,B_{xy,i}(b)}{\alpha(b)}} ~.
\end{align}
The flow in band $i$ stops if either the running gap reaches the running energy cutoff, $\Delta_{i+}(b) = u_{i+}/\alpha(b)$, or if the short distance cutoff equals the wire length, $\alpha(b) = L$, or if the finite temperature cuts off the divergences at $\alpha(b) = u_{+i}/T$. The first criterion defines the RG stage $b^*$ as $1= K_{i}\,\alpha(b^*)\,B_{xy,i}(b^*)/u_{i+}$, and thus

\begin{align}
b^* = \left(\frac{u_{i+}}{K_{i}\,\alpha\,B_{xy,i}}\right)^{1/(2-g_{i})}~.
\end{align}
The physical gap $\Delta_{i+}^* = \Delta_{i+}(b^*)$ in the electronic spectrum, provided that the RG flow is stopped by this first criterion, reads
\begin{align}
\Delta_{i+}^*= K_{i}\,B_{xy,i}~\left(\frac{u_{i+}}{K_{i}\,\alpha\,B_{xy,i}}\right)^{(1-g_{i})/(2-g_{i})}~.
\end{align}
If however the flow stops at $b^*_{L,T} = \alpha^*/\alpha$ with $\alpha(b^ {*}_{L,T}) = \alpha^{*} = {\rm min}\{L,u_{i+}/T\}$ due to the finite system size or temperature, the expansion of the cosine to second order yields a gap smaller than the finite size gap (or the temperature) $u_{i+}/\alpha^*$, and fluctuations beyond Gaussian order may be important for a given physical observable. Nevertheless using the Gaussian approximation, the Overhauser field induced gap can be expressed as

\begin{align}
\Delta_{i+}^* &= \frac{u_{i+}}{\alpha^*}\, \sqrt{\frac{K_{i}\,\alpha(b^*_{L,T})\,B_{xy,i}(b^*_{L,T})}{ u_{i+}}}\text{ .}
\end{align}
The renormalized Overhauser fields at the end of the flow, $B_{xy,i}^* = B_{xy,i}({\rm{min}}\{b^*,b_{L,T}^*\})$, can be interpreted as renormalized hyperfine coupling $A_i^* = A_0 \, B_{xy,i}^*/B_{xy,i}$ for the $\phi_{i+}$ modes (we recall that $B_{xy,i}\sim A_0$). This renormalized hyperfine coupling expresses how the ordered electronic modes collectively couple to the the polarized nuclear spins.

\subsection{Residual nuclear spin Hamiltonian}\label{subsec:residual_rkky}
Assuming the presence of two nuclear spin helices of the form $\eqref{eq:actual_overhauser_field}$ at low temperatures, we found that the fields $\phi_{1+}\sim \phi_{1c}+\theta_{1s}$ and $\phi_{2+}\sim \phi_{2c}+\theta_{2s}$ are gapped. Rather than through an exchange interaction mediated by all electronic modes, the nuclear spins thus feel the electrons mainly through an RKKY interaction mediated by the remaining gapless $\phi_{i-}$ modes, as well as through the finite magnetization of the electron spins of the $\phi_{i+}$ modes, which form helical spin density waves locked to the nuclear spin helices.\cite{braunecker_prb_09} Only outside the gapped momentum range, the $\phi_{i+}$ modes contribute to the RKKY exchange, which is then again of the high temperature form discussed in Sec.~\ref{subsec:high_t}. The dominant RKKY coupling $J_{\rm s}$ therefore exhibits two additional dips on the shoulders of each large dip signaling the onset of the exchange mediated by the $\phi_{i+}$ modes (the exchange mediated by the $\phi_{i+}$ modes is thus similar to the polarization function discussed in Ref.~\onlinecite{Badalyan_10}). The presence of the large nearby central dip, however, renders these side dips negligible.

\begin{figure}
\centering
\includegraphics[scale=1.0]{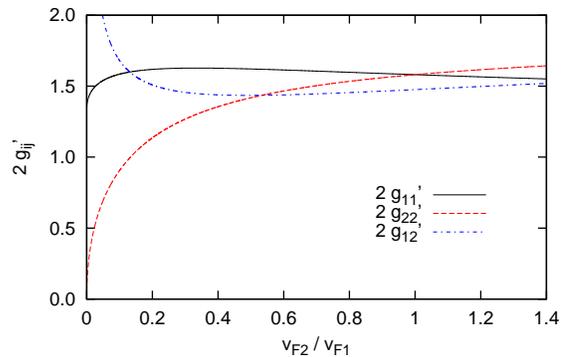}
\caption{Exponents $2 g_{ij}'$ of the residual RKKY exchange couplings after the formation of nuclear spin helices at momenta $2 k_{F1}$ and $2k_{F2}$, as a function of $v_{F2}/v_{F1}$ for fixed $v_{F1}$ and interaction strength $U$ such that $K_{1c} = 0.5$.}
\label{fig:scaling_exponents_small_res_self}
\end{figure}

To make our calculation self-consistent, we thus have to recalculate the RKKY interaction mediated only by the $\phi_{i-}$ modes. As detailed in Appendix \ref{append:self_cons_rkky}, we obtain the residual exchange couplings in $x$ and $y$ direction as

\begin{subequations}
\begin{align}\label{eq:spin_suscept_two_bands}
 J_{{\rm s}, q}^{xx}{}' &= J_{{\rm s}, q}^{yy}{}' = J_{{\rm s}, q, 1}' + J_{{\rm s}, q, 2}' ~,\\
  J_{{\rm a}, q}^{xx}{}' &= J_{{\rm a}, q}^{yy}{}' = J_{{\rm a}, q, 12}' + J_{{\rm a}, q, 21}' ~.
\end{align}
\end{subequations}
At finite temperatures, the RKKY exchange couplings in the symmetric sector read

\begin{align}
J_{{\rm s}, q, i}' &
 =-\frac{A_0^2\,a\,\sin\left(\pi g_{ii}'\right)}{32\pi^2 u_{i-}}\,\left(\frac{\beta u_{i-}}{2 \pi \alpha}\right)^{2-2g_{ii}'}\\
 &\times\sum_{\kappa=\pm}\Gamma^2\left(1-g_{ii}'\right)\,\left|\frac{\Gamma\left(\frac{g_{ii}'}{2}-i \frac{\beta u_{i-}}{4 \pi}(q+\kappa 2 k_{Fi})\right)}{\Gamma\left(\frac{2-g_{ii}'}{2}-i \frac{\beta u_i'}{4 \pi}(q+\kappa 2 k_{Fi})\right)}\right|^2\nonumber
\end{align}
where $\Gamma$ is the standard Gamma function, $u_{i-} = (u_{ic}/K_{is}+ u_{is}\,K_{ic})/(K_{ic}+1/K_{is})$ and where the exponents $g_{ii}'$ are given in Fig.~\ref{fig:scaling_exponents_small_res_self}. The intraband couplings have analogous expressions. At zero temperature, this translates to the divergent behavior

\begin{subequations}\label{eq:rkky_scaling_exponents_fp_d}
\begin{align}
 &\left.J_{{\rm s},q}^{xx}{}'\right|_{q\approx \pm 2k_{F1}} \sim \left|q\mp 2 k_{F1}\right|^{2g_{11}'-2} ~,\\
 &\left.J_{{\rm s},q}^{xx}{}'\right|_{q\approx \pm 2k_{F2}} \sim \left|q\mp 2 k_{F1}\right|^{2g_{22}'-2}~,\\
 &\left.J_{{\rm a},q}^{xx}{}'\right|_{q\approx \pm (k_{F1}+k_{F2})} \sim \left|q\mp (k_{F1}+k_{F2})\right|^{2g_{12}'-2}~.
\end{align}
\end{subequations}
 Note that for the calculation of the new exponents $g_{ij}'$, we have neglected renormalizations of the Luttinger liquid parameters and velocities within each band due to interband interactions during the flow between the fixed points $a$ and $c$ in Fig.~\ref{fig:rg_fixed_points}. Since this flow is rather short (we recall that the Overhauser fields are strongly RG relevant), and since interband interactions have a small effect unless $v_{F1}\approx v_{F2}$, we expect this approximation not to change our results.
 
Like for the single band wire, the RKKY interaction is now anisotropic. Correcting an error in Ref.~[\onlinecite{braunecker_prb_09}], the RKKY coupling along $z$ remains non-singular in the helically ordered state. This is an can be understood from Fig.~\ref{fig:scattering_gaps}. To obtain a divergent contribution to the $z$ component would require backscattering from a left-mover to a right-mover without spin flip. Such a process is however not possible in the helically ordered state where the right moving spin down and the left moving spin up particles are gapped. Technically, the spin susceptibility in $z$ direction is regularized by the fact that it involves the field $\theta_+$ canonically conjugate to the ordered field $\phi_+$, which cuts off its divergence.\cite{starykh_gapped_99,braunecker_prb_12,klinovaja_armchair_12,meng_loss_spin_susceptibility_sll_13}

To conclude, we find that the effective Hamiltonian for the nuclear spins including the feedback between the latter and the electrons reads

\begin{align}
H_{\rm RKKY}' &= \frac{1}{N} \sum_{q,\alpha}\Bigl(I_{{\rm s},-q}^\alpha \,\frac{J_{{\rm s},q}^{\alpha\alpha}{}'}{N_\perp^2}\, I_{{\rm s},q}^\alpha+ I_{{\rm a},-q}^\alpha \,\frac{J_{{\rm s},a}^{\alpha\alpha}{}'}{N_\perp^2}\,\, I_{{\rm a},q}^\alpha\Bigr)\label{eq:H_rkky_2}\\
&+\frac{1}{N_\perp} \,\sum_{i_\parallel}\,\left(A_1^*\,\langle\boldsymbol{S}_{11,  i_\parallel}\rangle+A_2^*\,\langle\boldsymbol{S}_{22,  i_\parallel}\rangle\right) \cdot \boldsymbol{I}_{{\rm s}, i_\parallel}\nonumber
\end{align}
As has been discussed in the single subband case, the effective magnetic fields $\sim \langle\boldsymbol{S}_{jj,  i_\parallel}\rangle$ acting on the nuclear spins due to the polarization of the (partially) ordered electron system are however negligible compared to the residual RKKY interaction and will therefore be omitted in the following.\cite{braunecker_prb_09}

\section{Coexistence of helices at $2k_{F1}$ and $2 k_{F2}$ and onset of nuclear spin order at finite temperature}\label{sec:order_and_finite_t}
Having established the effective theory in the presence of both hyperfine coupling and interband interactions, we are finally in the position to analyze the ground state of the system at low temperatures, where the nuclear spins order in a superposition of two helices at momenta $2k_{F1}$ and $2k_{F2}$ with magnetizations $m_{2k_{F1}}$ and $m_{2k_{F2}} = 1-m_{2k_{F1}}$ at zero temperature, see Eq.~\eqref{eq:actual_overhauser_field}. The coexistence of these two helices can be inferred from the fact that the energy loss of orienting parts of the nuclear spins into the shallower minimum of the RKKY interaction, see Fig.~\ref{fig:ja_js}, is overcompensated by the energy gain due to gapping out a second electronic sector. Considering for example $k_{F2} < k_{F1}$, the deepest minimum of the RKKY exchange is at momentum $2k_{F2}$. The loss in both nuclear spin energy and the electronic gap of the second band while populating the second minimum with a small magnetization $m_{2k_{F1}}$ scales linearly in $m_{2k_{F1}}$, while the gain in energy due to the gap in the first subband scales as $m_{2k_{F1}}^{1/(2-g_1)} > m_{2k_{F1}}$. A similar argument does not hold for the helical interband polarization $m_{k_{F1}+k_{F2}}$ because the latter would gap out the electronic spectrum and suppress the RKKY interaction altogether, see Sec.~\ref{sec:helix_order}.  The ground state magnetizations can then be obtained by minimizing the total energy with respect to $m_{2k_{F1}}$, which depends on the ratio of $v_{F2}$ to $v_{F1}$, as well as the temperature (through the RKKY interaction). The nuclear spin order is also stable if interband spin density interactions or current-current density interactions were non-negligible. Being quadratic in the bosonic fields, these interactions could be taken into account by a generalized transformation diagonalizing the interband interaction after the helix formation in Sec.~\ref{subsec:residual_rkky}, and would therefore not modify the presence or absence of electronic gaps.

At larger temperatures, thermal fluctuations decrease the nuclear spin polarizations $m_{2k_{F1}}$ and $m_{2k_{F2}}$, which also has to be fed back into the minimization procedure. When the temperature is eventually high enough, thermal fluctuations entirely suppress the nuclear spin order. Very roughly, this is expected to occur when the thermal energy is larger than the RKKY exchange at the helix momentum. An upper bound to the critical temperature is defined by the residual RKKY interactions that profit from the reduced exponents $g_{ij}'$ defined in the ordered phase. This yields the estimate $k_B \, T_{ci} \sim I^2 \,J_{{\rm s},i}^{xx}{}'$ for the transition temperature $T_{ci}$ in band $i$. Since the two helices at momenta $2k_{F1}$ and $2 k_{F2}$ are sustained by different minima of the RKKY exchange coupling with different depths, two distinct ordering temperatures can be observed. We derive these ordering temperatures in an approximation known as the Tyablikov decoupling.\cite{nolting_book, Tyablikov_additional} This method allows the self-consistent determination of the critical temperatures associated with the two helical nuclear spin orders in a mean field type approach. We find that the helical order in band $i$ sets in for temperatures below

\begin{align}
 T_{ci} \approx \frac{2}{3}\,I^2\,\left(\frac{a}{\pi}\,\int dq\,\frac{\tilde{\alpha}_i(q)}{\tilde{\alpha}_i(q)^2-\tilde{\beta}_i(q)^2}\right)^{-1}~,\label{eq:tci}
\end{align}
with
\begin{subequations}
 \begin{align}
  \tilde{\alpha}_i(q) &= \left[2 A_i(q=0)-A_i(q)-J_{{\rm s}, i}^z(q)\right]~,\\
\tilde{\beta}_i(q) &= \left[A_i(q)-J_{{\rm s}, i}^z(q)\right]~.
 \end{align}
\end{subequations}
with $ A_i(q) = J_{{\rm s},i}^{xx}{}'(q+2k_{Fi})/2+J_{{\rm s},i}^{xx}{}'(q-2k_{Fi})/2$ and $J_{{\rm s}, i}^z(q)$ being the contributions of band $i$ to the total spin susceptibility (see Eq.~\eqref{eq:spin_suscept_two_bands}). Details of the derivation of the critical temperatures $T_{ci}$ can be found in Appendix \ref{append:tyablikov}. The spin susceptibilities are dipped around $q = \pm 2 k_{Fi}$, see Fig.~\ref{fig:ja_js}, and we thus have $\tilde{\alpha}_i^2 - \tilde{\beta}_i^2 \sim q^2$ close to $q=0$. The integral in Eq.~\eqref{eq:tci} is therefore infrared (IR) divergent. This divergence is cut off by the finite size of the sample at the momentum $|q|\sim 2\pi/L$. For not too long quantum wires, the integral can be approximated by the implicit equation (note that $J_{{\rm s},i}^{xx}{}'$ has a power law temperature dependence)

\begin{align}\label{eq:tci_final}
 T_{ci} \approx \frac{2}{3}\,I^2\,J_{{\rm s},i}^{xx}{}'(2k_{Fi}) ~,
\end{align}
which confirms our order of magnitude guess. These critical temperatures $T_{ci}$ found with the Tyablikov decoupling are consistent with the result obtained in a magnon analysis.\cite{braunecker_prb_09,braunecker_magnon_comment} For longer wires, however, one needs to compare the critical temperature given in Eq.~\eqref{eq:tci_final}, resulting from the non-singular part of the integral in Eq.~\eqref{eq:tci}, to the IR divergent contribution $\sim [(\partial^2 \tilde{\alpha}/\partial q^2)_{\rm dip}]^{-1}\,L a/\pi^2$. For $L\to\infty$, the divergent contribution dominates, and the critical temperature goes to zero. This limiting behavior is consistent with the general theorem that there is no order possible in these systems at finite temperatures in the thermodynamic limit.\cite{loss_mermin_wagner_11}

We thus find that the nuclear spins in a finite size quantum wire with two subbands form two distinct helices of momenta $2k_{F1}$ and $2k_{F2}$, and that these two orders have different critical temperatures. The superposition of helices with different pitch lengths results in a beating pattern for the nuclear spin polarization. Since each helix order gaps out one electronic mode, the quantum wire exhibits a two step reduction of the conductance from $4e^2/h$ to (approximately) $3e^2/h$ at the first ordering temperature, and to (approximately) $2e^2/h$ at the second ordering, which provides an experimental signature of the double helix order.

\section{Intra band vs.~interband order}\label{sec:inter_intra}
A number of consistency checks of the results obtained here can be directly translated from the single band case, and we refer the reader to Ref.~[\onlinecite{braunecker_prb_09}] for details. What requires a little more care is the stability of the order in the symmetric channel with respect the order in the antisymmetric channel. The assumption of an ordering in the symmetric channel was based on the more divergent spin susceptibility of the latter with respect to the disordered fixed point $b$ in Fig.~\ref{fig:rg_fixed_points}, or alternatively the higher RG relevance of the Overhauser field associated with this channel close to this fixed point (note that close to the fixed point $a$ of Fig.~\ref{fig:rg_fixed_points}, interband scattering processes are even irrelevant in the RG sense). If the ratio of Fermi velocities becomes of order one, however, the two orders  have comparable scaling dimensions. One can suspect that in this regime, also the electronic gaps due to either possible ordering, as well as the nuclear spin energies (set by the RKKY interactions and thus also controlled by the exponents $g_{ij}$ that determine the scaling dimensions of the Overhauser fields) will become of the same order, and that the ground state can only be inferred from a detailed comparison of the various possible nuclear spin orders and associated electronic states. Similar considerations apply to the limit of weakly interacting electrons, where all $g_{ij} \to 1$. For smaller $v_{F2}/v_{F1}$ and experimentally observed strongly interacting Luttinger liquids, however, a helical order of the nuclear spins in the antisymmetric channel can be excluded. 

For $v_{F2} \lesssim 0.5 \, v_{F1}$, specifically, the stronger residual spin susceptibility in the symmetric channel, see Fig.~\ref{fig:scaling_exponents_small_res_self}, ensures a helical ordering at $2k_{F1}$ and $2k_{F2}$ at low enough temperatures. For larger $v_{F2}/v_{F1}$, our analysis becomes inconsistent for $T\to 0$ when the divergent spin susceptibilities overrule any finite gap in the electronic sectors. Given however that the maximal ratio of Fermi velocities before a third band is filled is $v_{F2}/v_{F1}\sim 0.7$ (for two subbands in a single wire), most of the two subband regime will exhibit the formation of two helices at $2k_{F1}$ and $2k_{F2}$, while the regime $v_{F2} \to v_{F1}$ deserves further analysis elsewhere.

\section{Summary}
\begin{figure}
\centering
$(a)$\quad\includegraphics[scale=0.1]{nuclear_lattice}~~\raisebox{1cm}{+}~~\includegraphics[scale=0.1]{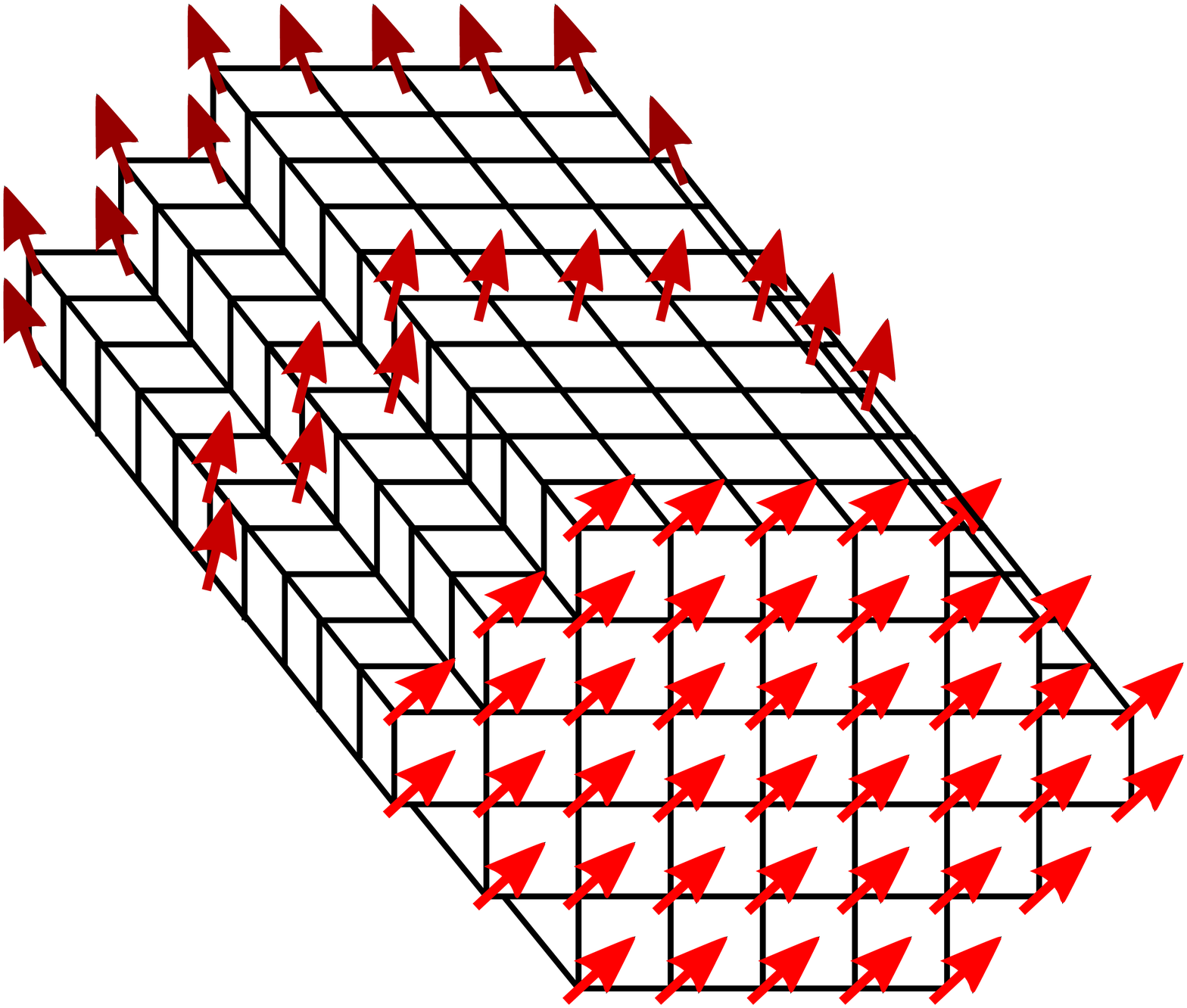}\\[0.75cm]
$(b)$\quad\includegraphics[scale=0.6]{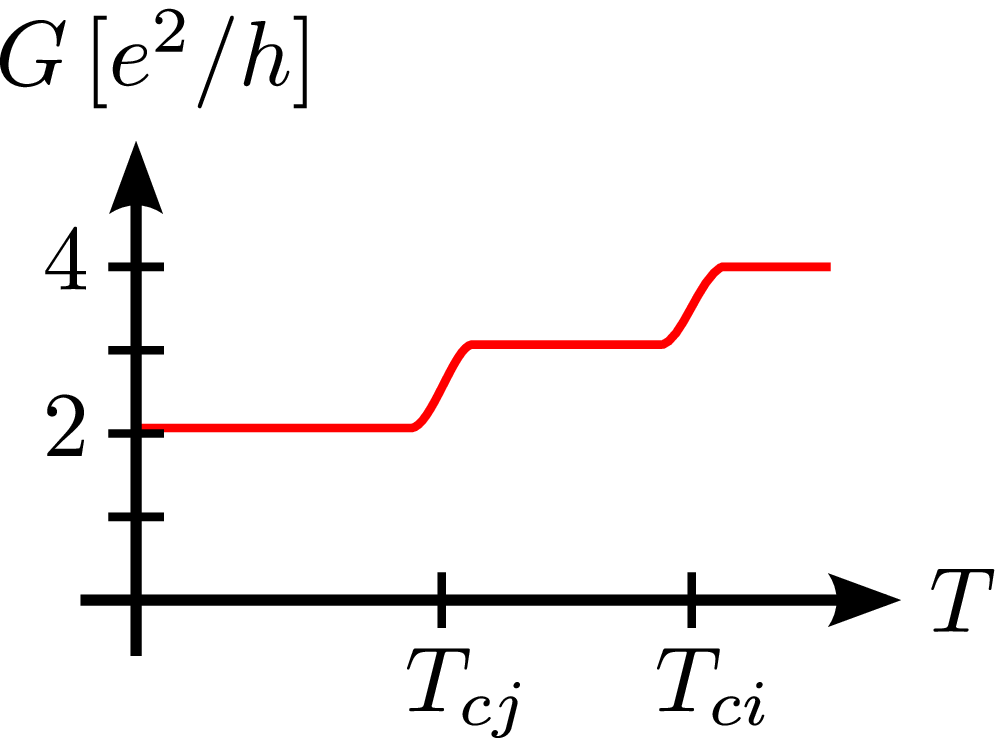}
\caption{Experimental signatures of the nuclear spin order. Panel $(a)$ shows the two superimposed helices of different pitch lengths $\lambda_1 = \pi/k_{F1}$ and $\lambda_2 = \pi/k_{F2}$ formed by the nuclear spins, leading to a beating pattern of the nuclear magnetization in real space. Panel $(b)$ depicts the conductance of the two subband quantum wire as a function of temperature. At the higher critical temperature $T_{ci}>T_{cj}$, the helix at momentum $2k_{Fi}$ forms, and the conductance drops from $4\, e^2/h$ to (approximately) $3\,e^2/h$. At the second ordering temperature $T_{cj}$, the second helix forms and further reduces the conductance to (approximately) $2\, e^2/h$.}
\label{fig:conductance}
\end{figure}

In this work, we have considered the interplay of electrons and nuclear spins in a two-subband quantum wire. Similar to a single subband wire, the hyperfine coupling between electrons and nuclear spins leads to an ordering of the nuclear spins and consequently a gap for half of the electronic spectrum.\cite{braunecker_prl_09,braunecker_prb_09} This gap is strongly renormalized by interaction effects. Different from the single subband case, the nuclear spins in a two-subband quantum wire form a superposition of two helices with distinct pitches $\lambda_1 = \pi/k_{F1}$ and $\lambda_2 = \pi/k_{F2}$, see Fig.~\ref{fig:conductance}$(a)$, which gives rise to a beating pattern in real space. The two helical orders set in at different temperatures. As a result, the conductance of the wire exhibits a stepwise reduction from $4\, e^2/h$ at highest temperatures to (approximately) $3\,e^2/h$ after the formation of the first helix, and finally to (approximately) $2\,e^2/h$ at the formation of the second helix. The behavior of the conductance, which provides an experimental signature of the double helical order in two-subband quantum wires, is depicted in Fig.~\ref{fig:conductance}$(b)$.

To establish these results, we have analyzed the quantum wire in a self-consistent analysis based on the well-separated dynamics of electrons and nuclear spins. As has been shown in Sec.~\ref{sec:model}, the hyperfine coupling between electrons and nuclear spins can be decomposed into a symmetric and an antisymmetric channel. The electrons thus mediate two kinds of RKKY interactions between the nuclear spins. As has been discussed in Sec.~\ref{sec:helix_order}, these different interaction channels compete for the formation of nuclear spin order. The final ground state has then been derived in a self-consistent approach that takes into account the interplay of nuclear spins and electrons by describing the nuclear spins as an effectively static Overhauser field for the electrons, while the latter mediate an effectively instantaneous RKKY interaction for the former. This calculation has been carried out in Sec.~\ref{sec:feedback}. We find that the two-subband quantum wire preferably orders in the symmetric hyperfine channel, which leads to the formation of the two superimposed nuclear spin helices of pitches $\lambda_1$ and $\lambda_2$ at lowest temperatures. At higher temperatures, each helix is associated with its own critical temperature, implying the above described reduction of the conductance in two steps as the temperature is lowered, see Sec.~\ref{sec:order_and_finite_t}.

\acknowledgements
We thank Pascal Simon, Bernd Braunecker, and Jelena Klinovaja for helpful discussions. This work has been supported by SNF, NCCR Nano, and NCCR QSIT.

\appendix

\section{RKKY interaction}
\label{append:rkky}
In order to derive the RKKY interaction by integrating out the fermionic degrees of freedom, we first rewrite the hyperfine coupling in momentum space and imaginary time. Using $I^\alpha_{(\cdot)i_\parallel}(\tau) = \sum_i e^{-iq z_i} I_{(\cdot)q}^\alpha(\tau)$ with $z_i = a \, i_\parallel$, where $i_\parallel$ denotes the longitudinal nuclear spin lattice site index and with $a$ being the lattice constant, we obtain the action corresponding to $H_{\rm ne}$ as

\begin{align}
 \mathcal{S}_{\rm ne} &= \frac{A_0}{N_\perp}\frac{1}{N}\int d\tau\sum_{q} \left(\boldsymbol{S}_{11,  q}(\tau)+\boldsymbol{S}_{22,  q}(\tau)\right)\cdot \boldsymbol{I}_{{\rm s},q}(\tau)\\
&+\ \frac{A_0}{N_\perp} \frac{1}{N}\int d\tau\sum_{q} \left(\boldsymbol{S}_{12,  q}(\tau)+\boldsymbol{S}_{21, q}(\tau)\right)\cdot \boldsymbol{I}_{{\rm a}, q}(\tau) ~,\nonumber
\end{align}
where $N = L/a$ is the number of lattice sites in longitudinal direction, and with $\boldsymbol{S}_{ij, q}(\tau)= \sum_{\alpha,\beta, k} c_{i\alpha k+q}^\dagger(\tau)(\boldsymbol{\sigma}_{\alpha\beta}/2) c_{j\alpha k}(\tau)^\pdag$. The RKKY interaction can be obtained by expanding the total action $\mathcal{S} = \mathcal{S}_{\rm e} + \mathcal{S}_{\rm ne}$ to second order in $\mathcal{S}_{\rm ne}$ and integrating over the fermions. Re-exponentiation then yields the action

\begin{align}
\mathcal{S}_{\rm RKKY} &= \frac{A_0^2}{N_{\perp}^2}\frac{1}{2N^2}\int d\tau \int d\tau' \sum_{q}\\
&\times\Bigl(I_{{\rm s},-q}^\alpha(\tau) \,\chi_{{\rm s},q}^{\alpha\beta}(\tau-\tau')\, I_{{\rm s},q}^\beta(\tau')\nonumber\\
&+ I_{{\rm a},-q}^\alpha(\tau) \,\chi_{{\rm a},q}^{\alpha\beta}(\tau-\tau')\, I_{{\rm a},q}^\beta(\tau')\Bigr)\nonumber
\end{align}
with the imaginary time spin susceptibilities

\begin{align}
 \chi_{\rm s,q}^{\alpha \beta}(\tau-\tau') &= \langle S_{11,-q}^\alpha(\tau) S_{11,q}^\beta(\tau') + S_{22,-q}^\alpha(\tau) S_{22,q}^\beta(\tau')  \rangle\label{eq:chi_s}~,\\
  \chi_{\rm a,q}^{\alpha \beta}(\tau-\tau') &= \langle S_{12,-q}^\alpha(\tau) S_{21,q}^\beta(\tau') + S_{21,-q}^\alpha(\tau) S_{12,q}^\beta(\tau')  \rangle ~.\label{eq:chi_a}
\end{align}
The averages are taken with respect to $\mathcal{S}_{\rm e}$ and evaluated in Appendix \ref{Appendix:suscept}. Given that the nuclear spin dynamics are associated with much longer time scales than the electron dynamics, we can approximate (with $\beta = 1/T$  in units of $k_B = 1$)

\begin{align}
\chi_{(\cdot),q}(\tau-\tau') &= \frac{1}{\beta}\sum_{\omega_n}e^{-i\omega_n(\tau-\tau')}\chi_{(\cdot),q}(\omega_n)\\
&\approx \frac{1}{\beta}\sum_{\omega_n}e^{-i\omega_n(\tau-\tau')}\chi_{(\cdot),q}(\omega_n=0) \nonumber\\
&= \delta(\tau-\tau') \chi_{(\cdot),q}(\omega_n=0) ~.\nonumber
\end{align}
The real time Hamiltonian corresponding to the action $\mathcal{S}_{\rm RKKY}$ can finally be found by analytical continuation to the retarded spin susceptibilities,
\begin{align}
H_{\rm RKKY} &= \frac{A_0^2}{N_{\perp}^2}\frac{1}{2N^2} \sum_{q}\\
&\times\Bigl(I_{{\rm s},-q}^\alpha(t) \,\chi_{{\rm s},q}^{\alpha\beta,\rm ret}(\omega\to0)\, I_{{\rm s},q}^\beta(t)\nonumber\\
&+ I_{{\rm a},-q}^\alpha(t) \,\chi_{{\rm a},q}^{\alpha\beta,\rm ret}(\omega \to 0)\, I_{{\rm a},q}^\beta(t)\Bigr)\nonumber
\end{align}

\section{Diagonalization of the electron sector}\label{append:diag}
In order to diagonalize the electronic Hamiltonian $H_{\rm e}$, we first bosonize the annihilation operator for a particle of spin $\sigma$ as

\begin{align}
r_{j,\sigma}(z) = \frac{U_{rj\sigma}}{\sqrt{2\pi\alpha}}\,e^{-i(r\phi_{j\sigma}(z)-\theta_{j\sigma}(z))} ~,
\end{align}
where $r=R,L\equiv+,-$, the corresponding Klein factors are denoted as $U_{rj\sigma}$, and $\alpha$ is a short distance cutoff (here taken to be the lattice spacing). The bosonic fields $\phi_{j\sigma}$ and $\theta_{j\sigma}$ fulfill the standard commutation relation $[\phi_{j\sigma}(z),\theta_{j'\sigma'}(z')]=\delta_{jj'}\delta_{\sigma\sigma'}\,(i\pi/2)\,\text{sgn}(z'-z)$.\cite{giamarchi_book} Since the interaction $U$ in Eq.~\eqref{eq:total_hamilton_1} is of density-density type, it can be taken into account by an exact basis transformation. Because furthermore the Fermi velocities $v_{F1}$ and $v_{F2}$ in the two bands are different, it is most convenient to first diagonalize each subband separately by introducing spin and charge degrees of freedom via the canonical transformation $\phi_{j\stackrel{c}{s}}(z) = (\phi_{j\uparrow}\pm\phi_{j\downarrow})/\sqrt{2}$ and $\theta_{j\stackrel{c}{s}}(z) = (\theta_{j\uparrow}\pm\theta_{j\downarrow})/\sqrt{2}$. In this basis, the electronic Hamiltonian reads

\begin{align}
 H_{\rm e} &= \sum_{i=1,2}\sum_{j=c,s}\int \frac{dz}{2\pi}\,\left(\frac{u_{ij}}{K_{ij}}\,(\partial_z \phi_{ij})^2 + u_{ij} K_{ij} \, (\partial_z\theta_{ij})^2\right)\nonumber\\
 &+ \int \frac{dz}{2\pi}\,\frac{4U}{\pi}\, (\partial_z \phi_{1c})(\partial_z \phi_{2c}),
\end{align}
where the effective velocities  and Luttinger parameters are as usual given by

\begin{subequations}
\begin{align}
 K_{ic} &= \left(\sqrt{1+\frac{2 U}{\pi v_{Fi}}}\right)^{-1} ~,~~u_{ic} = \frac{v_{Fi}}{K_{ic}} ~,\\
 K_{is} &=1 ~,~~u_{is} = v_{Fi} ~.
\end{align}
\end{subequations}
The diagonal electronic Hamiltonian is now obtained by a second canonical transformation

\begin{subequations}\label{eq:diagonalization_trafo}
 \begin{align}
\phi_{1c}&= \sqrt{\frac{v_{F1}}{u_{c+}\,(1+\mathcal{A}_c^2)}}\,\phi_{c+}+\sqrt{\frac{\mathcal{A}_c^2\,v_{F1}}{u_{c-}(1+\mathcal{A}_c^2)}}\,\phi_{c-}~,\label{eq:diagonalization_trafo_pc1}\\
\phi_{2c}&= \sqrt{\frac{\mathcal{A}_c^2\,v_{F2}}{u_{c+}(1+\mathcal{A}_c^2)}}\,\phi_{c+}- \sqrt{\frac{v_{F2}}{u_{c-}(1+\mathcal{A}_c^2)}}\,\phi_{c-}~,\\
\theta_{1c}&=\sqrt{\frac{u_{c+}}{v_{F1}(1+\mathcal{A}_c^2)}}\,\theta_{c+} + \sqrt{\frac{\mathcal{A}_c^2\,u_{c-}}{v_{F1}(1+\mathcal{A}_c^2)}}\,\theta_{c-}~, \\
\theta_{2c}&=\sqrt{\frac{\mathcal{A}_c^2\,u_{c+}}{v_{F2}(1+\mathcal{A}_c^2)}}\,\theta_{c+} - \sqrt{\frac{u_{c-}}{v_{F2}(1+\mathcal{A}_c^2)}}\,\theta_{c-} ~.
 \end{align}
\end{subequations}
with the velocities

\begin{align}\label{eq:mixed_velocities}
 u_{c\pm} = \sqrt{\frac{u_{1c}^2+u_{2c}^2}{2}\pm\sqrt{\left(\frac{u_{1c}^2-u_{2c}^2}{2}\right)^2+\left(\frac{2U}{\pi}\right)^2 v_{F1} v_{F2}}}
\end{align}
and with
\begin{align}\label{eq:mixing_coeff_coulomb}
 \mathcal{A}_c= \frac{(4U/\pi)\,\sqrt{v_{F1}v_{F2}}}{\sqrt{\left(u_{1c}^2-u_{2c}^2\right)^2+\left(4U/\pi\right)^2\,v_{F1} v_{F2}}+u_{1c}^2-u_{2c}^2} ~.
\end{align}
The electronic Hamiltonian can then be written as

\begin{align}\label{eq:diag_el_ham}
 H_{\rm e} &= \sum_{k=\pm}\frac{u_{ck}}{2\pi}\int dz \,\left( \left(\partial_z\phi_{ck}\right)^2 + \left(\partial_z\theta_{ck}\right)^2\right)\\
  &+\sum_{j=1,2}\frac{v_{Fj}}{2\pi}\int dz \,\left( \left(\partial_z\phi_{sj}\right)^2 + \left(\partial_z\theta_{sj}\right)^2\right).\nonumber
\end{align}
We note that a generalized form of the transformation given in Eq.~\eqref{eq:diagonalization_trafo} would also allow to take into account the spin density-density interaction, charge current-current interaction and spin current-current interaction neglected here.

\section{Evaluation of the spin susceptibilities}
\label{Appendix:suscept}
The real time RKKY Hamiltonian given in Eq.~\eqref{eq:H_rkky} depends on the zero frequency components of the retarded spin susceptibilities. The latter can most conveniently be derived starting from the imaginary time susceptibilities given in Eqs.~\eqref{eq:chi_s} and \eqref{eq:chi_a}. We obtain

\begin{align}
 &\langle S_{ij,-q}^\alpha(\tau) S_{i'j',q}^\beta(\tau') \rangle = \sum_{k,k'}\sum_{a,b,a',b'} \frac{\sigma_{ab}^\alpha}{2} \frac{\sigma^\beta_{a'b'}}{2}\nonumber\\
 &\times \langle c_{ia k-q}^\dagger(\tau) c_{j b k}^\pdag(\tau) c_{i'a' k'+q}^\dagger(\tau') c_{j'b' k'}^\pdag(\tau') \rangle\\
 &= \sum_{a,b,a',b'} \frac{\sigma_{ab}^\alpha}{2} \frac{\sigma^\beta_{a'b'}}{2}\,L\, \int d(z-z') e^{i q (z-z')}\nonumber\\
 &\times\langle c_{ia}^\dagger(z,\tau) c_{j b}^\pdag(z,\tau) c_{i'a'}^\dagger(z',\tau') c_{j'b'}^\pdag(z',\tau') \rangle\nonumber ~.
\end{align}
Since the Hamiltonian conserves the total spin, these averages (evaluated with respect to $H_{\rm e}$) are diagonal in spin space. In addition, we find that forward scattering contributions have a non-singular spin susceptibility, while backscattering events result in divergences. We therefore only evaluate the backscattering terms. To this end, the electronic Hamiltonian is bosonized and (depending on the fixed point one is interested in) diagonalized as detailed in Appendix \ref{append:diag}. The divergent part of the $x$ and $y$ components of the spin susceptibility within the first band are identical and read

\begin{align}
 \chi_{11,q}^x &=  \chi_{11,q}^y = \sum_{r\sigma} \frac{L}{4} \,\int d(z-z') e^{i(q-2 r k_{F1})(z-z')}\\ 
 &\times\langle r_{1\sigma}^\dagger(z,\tau) \bar{r}_{1\bar{\sigma}}^\pdag(z,\tau) \bar{r}_{1\bar{\sigma}}^\dagger(z',\tau') r_{1\sigma}^\pdag(z',\tau')\rangle\nonumber\\
 &= -\frac{L}{4(2\pi\alpha)^2} \,\int d(z-z') e^{iq(z-z')}\,\Bigl(e^{-i2k_{F1}(z-z')}\nonumber\\ 
 &\times\Bigl[\langle e^{i\sqrt{2}(\phi_{1c}(z,\tau)+\theta_{1s}(z,\tau)-\phi_{1c}(z',\tau')-\theta_{1s}(z',\tau'))}\rangle\nonumber\\
 &+\langle e^{i\sqrt{2}(\phi_{1c}(z,\tau)-\theta_{1s}(z,\tau)-\phi_{1c}(z',\tau')+\theta_{1s}(z',\tau'))}\rangle\nonumber\Bigr]+\rm{h.c.}\Bigr)
\end{align}
with $r=R,L\equiv+,-$. The backscattering part of the $z$-component reads
\begin{align}
 \chi_{11,q}^z &= \sum_{r\sigma} \frac{L}{4} \, \int d(z-z')e^{i(q-2 r k_{F1})(z-z')}\label{eq:append_chi}\\ 
 &\times\langle r_{1\sigma}^\dagger(z,\tau) \bar{r}_{1\sigma}^\pdag(z,\tau) \bar{r}_{1\sigma}^\dagger(z',\tau') r_{1\sigma}^\pdag(z',\tau')\rangle\nonumber\\
 &= -\frac{L}{4(2\pi\alpha)^2} \,\int d(z-z')  e^{iq(z-z')}\,\Bigl(e^{-i2k_{F1}(z-z')}\nonumber\\ 
 &\times\Bigl[\langle e^{(i\sqrt{2}(\phi_{1c}(z,\tau)+\phi_{1s}(z,\tau)-\phi_{1c}(z',\tau')-\phi_{1s}(z',\tau'))}\rangle\nonumber\\
 &+\langle e^{i\sqrt{2}(\phi_{1c}(z,\tau)-\phi_{1s}(z,\tau)-\phi_{1c}(z',\tau')+\phi_{1s}(z',\tau'))}\rangle\nonumber\Bigr]+\rm{h.c.}\Bigr)
\end{align}
The spin susceptibility within the second band can be obtained from the one in the first band by exchanging the index $1\to2$. The spin susceptibility between the bands finally has the divergent part

\begin{align}
 \chi_{12,q}^x &=  \chi_{12,q}^y = \sum_{r\sigma} \frac{L}{4} \,\int d(z-z') e^{i(q-r (k_{F1}+k_{F2}))(z-z')}\nonumber\\ 
 &\times\langle r_{1\sigma}^\dagger(z,\tau) \bar{r}_{2\bar{\sigma}}^\pdag(z,\tau) \bar{r}_{2\bar{\sigma}}^\dagger(z',\tau') r_{1\sigma}^\pdag(z',\tau')\rangle\\
 &= \frac{-L}{4(2\pi\alpha)^2} \,\int d(z-z') e^{iq(z-z')}\Bigl(e^{-i (k_{F1}+k_{F2})(z-z')}\nonumber\\ 
 &\times\Bigl[\langle e^{i(1/\sqrt{2})(\phi_{1c}(z,\tau)+\phi_{1s}(z,\tau)+\phi_{2c}(z,\tau)-\phi_{2s}(z,\tau))}\nonumber\\
& \times e^{-i(1/\sqrt{2})(\theta_{1c}(z,\tau)+\theta_{1s}(z,\tau)-\theta_{2c}(z,\tau)+\theta_{2s}(z,\tau))}\nonumber\\ 
&\times e^{-i(1/\sqrt{2})(\phi_{1c}(z',\tau')+\phi_{1s}(z',\tau')+\phi_{2c}(z',\tau')-\phi_{2s}(z',\tau'))}\nonumber\\
& \times e^{i(1/\sqrt{2})(\theta_{1c}(z',\tau')+\theta_{1s}(z',\tau')-\theta_{2c}(z',\tau')+\theta_{2s}(z',\tau'))}\rangle\nonumber\\
 &+\langle e^{i(1/\sqrt{2})(\phi_{1c}(z,\tau)-\phi_{1s}(z,\tau)+\phi_{2c}(z,\tau)+\phi_{2s}(z,\tau))}\nonumber\\
& \times e^{-i(1/\sqrt{2})(\theta_{1c}(z,\tau)-\theta_{1s}(z,\tau)-\theta_{2c}(z,\tau)-\theta_{2s}(z,\tau))}\nonumber\\ 
&\times e^{-i(1/\sqrt{2})(\phi_{1c}(z',\tau')-\phi_{1s}(z',\tau')+\phi_{2c}(z',\tau')+\phi_{2s}(z',\tau'))}\nonumber\\
& \times e^{i(1/\sqrt{2})(\theta_{1c}(z',\tau')-\theta_{1s}(z',\tau')-\theta_{2c}(z',\tau')-\theta_{2s}(z',\tau'))}\rangle\Bigr]\nonumber\\
&+(k_{Fi}\to -k_{Fi},\phi_{ij}\to - \phi_{ij} ) \Bigr)~,\nonumber
\end{align}
In addition, there is a contribution $\chi_{21,q}^x = \chi_{21,q}^y$ which can be obtained from $\chi_{12,q}^x = \chi_{12,q}^y$ by the index swap $1\leftrightarrow2$. The $z$ component finally has the expression
\begin{align}
 \chi_{12,q}^z &= \sum_{r\sigma} \frac{L}{4} \, \int d(z-z') e^{i(q-r (k_{F1}+k_{F2}))(z-z')}\\ 
 &\times\langle r_{1\sigma}^\dagger(z,\tau) \bar{r}_{2\sigma}^\pdag(z,\tau) \bar{r}_{2\sigma}^\dagger(z',\tau') r_{1\sigma}^\pdag(z',\tau')\rangle\nonumber\\
 &= \frac{-L}{4(2\pi\alpha)^2} \, \int d(z-z') e^{iq(z-z')}\,\Bigl(e^{-i(k_{F1}+k_{F2})(z-z')}\nonumber\\ 
 &\times\Bigl[\langle e^{i(1/\sqrt{2})(\phi_{1c}(z,\tau)+\phi_{1s}(z,\tau)+\phi_{2c}(z,\tau)+\phi_{2s}(z,\tau))}\nonumber\\
& \times e^{-i(1/\sqrt{2})(\theta_{1c}(z,\tau)+\theta_{1s}(z,\tau)-\theta_{2c}(z,\tau)-\theta_{2s}(z,\tau))}\nonumber\\ 
&\times e^{-i(1/\sqrt{2})(\phi_{1c}(z',\tau')+\phi_{1s}(z',\tau')+\phi_{2c}(z',\tau')+\phi_{2s}(z',\tau'))}\nonumber\\
& \times e^{i(1/\sqrt{2})(\theta_{1c}(z',\tau')+\theta_{1s}(z',\tau')-\theta_{2c}(z',\tau')-\theta_{2s}(z',\tau'))}\rangle\nonumber\\
 &+\langle e^{i(1/\sqrt{2})(\phi_{1c}(z,\tau)-\phi_{1s}(z,\tau)+\phi_{2c}(z,\tau)-\phi_{2s}(z,\tau))}\nonumber\\
& \times e^{-i(1/\sqrt{2})(\theta_{1c}(z,\tau)-\theta_{1s}(z,\tau)-\theta_{2c}(z,\tau)+\theta_{2s}(z,\tau))}\nonumber\\ 
&\times e^{-i(1/\sqrt{2})(\phi_{1c}(z',\tau')-\phi_{1s}(z',\tau')+\phi_{2c}(z',\tau')-\phi_{2s}(z',\tau'))}\nonumber\\
& \times e^{i(1/\sqrt{2})(\theta_{1c}(z',\tau')-\theta_{1s}(z',\tau')-\theta_{2c}(z',\tau')+\theta_{2s}(z',\tau'))}\rangle\Bigr]\nonumber\\
&+(k_{Fi}\to -k_{Fi},\phi_{ij}\to - \phi_{ij} ) \Bigr) ~,\nonumber
\end{align}
plus again an additional contribution that is obtained under $1\leftrightarrow2$. These averages are now evaluated with respect to the diagonalized Hamiltonian $H_{\rm e}$ given in Eq.~\eqref{eq:diag_el_ham}. We explicitly state the calculation of $\chi_{11}^x$, the other susceptibilities can be evaluated analogously. Following Ref.~[\onlinecite{giamarchi_book}] and using Eq.~\eqref{eq:diagonalization_trafo_pc1}, we find that

\begin{align}
 \chi_{11,q}^x &=-\frac{L}{2(2\pi\alpha)^2} \,\int d(z-z')\\
 &\times \left(e^{i(q+2k_{F1})(z-z')}+e^{i(q-2k_{F1})(z-z')}\right)\nonumber\\ 
 &\times e^{-(A^2F_1^{(c+)}(z-z',\tau-\tau')+B^2F_1^{(c-)}(z-z',\tau-\tau'))}\nonumber\\
 &\times e^{-(1/K_{s1})F_1^{(s1)}(z-z',\tau-\tau')}\nonumber
\end{align}
with
\begin{align}
 A=\sqrt{\frac{v_{F1}}{u_{c+}\,(1+\mathcal{A}_c^2)}} ~,~~B=\sqrt{\frac{\mathcal{A}_c^2\,v_{F1}}{u_{c-}(1+\mathcal{A}_c^2)}} ~,
\end{align}
and where the function $F_1^{(\cdot)}(z,\tau)$ is at zero temperature given by

\begin{equation}
F_{1}^{(\cdot)}(z, \tau) =  \log\left(\frac{\sqrt{z^2+(u_{(\cdot)}|\tau|+\alpha)^2}}{\alpha}\right) ~,
\end{equation}
where $u_{(\cdot)}$ is the velocity associated with the respective mode. At finite temperatures, it becomes

\begin{equation}
F_{1}^{(\cdot)}(z, \tau) = 
\log\left(\frac{\beta u_{(\cdot)}}{\pi\alpha}\sqrt{\sinh^2\left(\frac{\pi z}{\beta u_{(\cdot)}}\right)+\sin^2\left(\frac{\pi \tau}{\beta}\right)}\right)
\end{equation}
(where $\beta = 1/T$ in units of $k_B = 1$). Being mainly interested in the scaling behavior of susceptibilities, i.e.~their power law exponents, one can neglect the difference in velocities and obtains

\begin{align}
  \chi_{11,q}^x &=-\frac{L}{2(2\pi\alpha)^2} \,\int dz \left(e^{i(q+2k_{F1})z}+e^{i(q-2k_{F1})z}\right)\nonumber\\ 
 &\times\left(\frac{\pi\alpha/(\beta u)}{\sqrt{\sinh^2\left(\frac{\pi z}{\beta u}\right)+\sin^2\left(\frac{\pi (\tau-\tau')}{\beta}\right)}}\right)^{A^2 + B^2 +  1/K_{1s}}\label{eq:scaling_chi_11_x_t0}
\end{align}
Here, $u$ denotes the common velocity of the different modes in this approximation. The physical response of the system can now be obtained by analytic continuation of Eq.~\eqref{eq:scaling_chi_11_x_t0}. The latter is described in detail in Ref.~[\onlinecite{giamarchi_book}] and yields
\begin{align}
 &\left.\chi_{11,q}^{x,\rm ret}\right|_{\omega\to0} =-\frac{L\,\sin\left(\pi g_{11}\right)}{8\pi^2 u}\,\left(\frac{\beta u}{2 \pi \alpha}\right)^{2-2g_{11}}\\
 &\times\sum_{\kappa=\pm}\Gamma^2\left(1-g_{11}\right)\,\left|\frac{\Gamma\left(\frac{g_{11}}{2}-i \frac{\beta u}{4 \pi}(q+\kappa 2 k_{F1})\right)}{\Gamma\left(\frac{2-g_{11}}{2}-i \frac{\beta u}{4 \pi}(q+\kappa 2 k_{F1})\right)}\right|^2\nonumber
\end{align}
where $\Gamma$ is the standard Gamma function and with

\begin{equation}
g_{11}= \frac{A^2 + B^2 +  1/K_{1s}}{2}~.
\end{equation}
At zero temperature, this translates into
\begin{align}
 \left.\chi_{11,q}^{x,\rm ret}\right|_{\omega\to0} &=-\frac{L\,\sin\left(\pi g_{11}\right)}{8\pi^2 u}\,\sum_{\kappa=\pm}\left|\frac{2}{\alpha(q+\kappa 2 k_{F1})}\right|^{2-2g_{11}} ~.
\end{align}
The spin susceptibility within the first band $\chi_{11}$ thus diverges with a power law exponent $2-2g_{11}$ at the backscattering wave vector $2k_{F1}$. We find similar power law divergences for $\chi_{22}$, $\chi_{21}$, and $\chi_{12}$ characterized by exponents $2-2g_{ij}$. These exponents are plotted in Figs.~\ref{fig:scaling_exponents_small}  and \ref{fig:scaling_exponents_large}. Due to the transformation detailed in Eq.~\eqref{eq:diagonalization_trafo}, they depend on the interaction strength between the two bands as well as the ratio of Fermi velocities $v_{F1}/v_{F2}$.

\section{Overhauser fields and their coupling to the electronic modes}\label{append:overhauser_fields}
In this Appendix, we shortly review how an Overhauser field of one of the two effective nuclear spins affects the electrons in the quantum wire. Let us first start with a helical field of positive helicity in the symmetric channel at $2k_{F1}$,

\begin{align}
\boldsymbol{B}_{\rm Ov, \rm s}(z) = \langle \boldsymbol{I}_{{\rm s}, z}\rangle = B_0 \, \begin{pmatrix}\cos(2k_{F1}z)\\\sin(2k_{F1}z)\\0\end{pmatrix} ~.
\end{align}
This Overhauser field is coupled to the electrons by the hyperfine interaction,

\begin{align}
H_{\rm en} &= \frac{A_0}{N_\perp} \,\int dz \, \left(\boldsymbol{S}_{11}(z)+\boldsymbol{S}_{22}(z)\right)\cdot \boldsymbol{B}_{\rm Ov, \rm s}(z)\\
&=\frac{A_0\,B_0}{2N_\perp} \int dz\,\left(c_{1\uparrow}^\dagger \, e^{-i2k_{F1}z}\, c_{1\downarrow}^\pdag + \text{h.c.}\right)\nonumber\\
&+\frac{A_0\,B_0}{2N_\perp} \int dz\,\left(c_{2\uparrow}^\dagger \, e^{-i2k_{F1}z}\, c_{2\downarrow}^\pdag + \text{h.c.}\right)\text{ .}\nonumber
\end{align}
The only non-oscillating cosine-terms deriving from this interaction (unless the two Fermi wave vectors are commensurate) are backscattering terms of a right-moving spin down and a left moving spin up within the first band. Dropping the Klein factors that cancel out when taking averages subsequently, the bosonized version of this term reads

\begin{align}
H_{\rm en}^{\rm non-osc.} = \frac{A_0 B_0}{2 \pi \alpha N_{\perp}} \int dz\,\cos\left(\sqrt{2}(\phi_{1c} + \theta_{1s})\right) \text{ .}
\end{align}
This term is strongly RG relevant and gaps out the combination $\phi_{c1} + \theta_{s1}$ in the first band, corresponding to right moving spin down particles and left moving spin up particles (see Ref.~[\onlinecite{braunecker_prb_09}] for a detailed discussion). An Overhauser field of negative helicity would gap out the combination $\phi_{c1} - \theta_{s1}$ that corresponds to right moving spin down particles and left moving spin up particles. A helical Overhauser field with momentum $2k_{F2}$, on the other hand, would gap out the corresponding combinations $\phi_{c2}\pm\theta_{c2}$ in the second band.

We now turn to an Overhauser field in the antisymmetric channel. For

\begin{align}
\boldsymbol{B}_{\rm Ov, \rm a}(z) = \langle \boldsymbol{I}_{{\rm a}, z}\rangle = B_0 \, \begin{pmatrix}\cos((k_{F1}+k_{F2})z)\\-\sin((k_{F1}+k_{f2})z)\\0\end{pmatrix} ~,
\end{align}
the hyperfine interaction is

\begin{align}
H_{\rm en} &= \frac{A_0}{N_\perp} \,\int dz \, \left(\boldsymbol{S}_{12}(z)+\boldsymbol{S}_{21}(z)\right)\cdot \boldsymbol{B}_{\rm Ov, \rm a}(z)\\
&=\frac{A_0\,B_0}{2N_\perp} \int dz\,\left(c_{1\uparrow}^\dagger \, e^{i(k_{F1}+k_{F2})z}\, c_{2\downarrow}^\pdag + \text{h.c.}\right)\nonumber\\
&+\frac{A_0\,B_0}{2N_\perp} \int dz\,\left(c_{2\uparrow}^\dagger \, e^{i(k_{F1}+k_{F2})z}\, c_{1\downarrow}^\pdag + \text{h.c.}\right)\text{ .}\nonumber
\end{align}
Bosonizing this Hamiltonian, we find two non-oscillating cosine-terms. The first line allows for backscattering between right moving spin up particles in the second band and left moving spin down particles in the first band, while the second one allows for scattering between right moving spin up particles in the first band and left moving spin down particles in the second band. Bosonization finally yields

\begin{align}
H_{\rm en}^{\rm non-osc.} &= \frac{A_0 B_0}{2 \pi \alpha N_{\perp}} \int dz\,\cos\left(\sqrt{2}(\phi_{12c}-\theta_{12s}))\right) \\
&+ \frac{A_0 B_0}{2 \pi \alpha N_{\perp}} \int dz\,\cos\left(\sqrt{2}(\phi_{21c}-\theta_{21s}))\right)~,\nonumber
\end{align}
where

\begin{align}
\phi_{ijc} = \frac{\phi_{i\uparrow} + \phi_{j\downarrow}}{\sqrt{2}}\quad,\quad\theta_{ijs} = \frac{\theta_{i\uparrow} - \theta_{j\downarrow}}{\sqrt{2}}~.
\end{align}
Like the cosine-terms within a single band, the interband cosine-terms are RG-relevant if the exponent $2g_{12}$ defined in Eq.~\eqref{eq:rkky_scaling_exponents}
 is smaller than 2 and open up gaps for $\phi_{12c} - \theta_{12s}$ and $\phi_{21c} - \theta_{21s}$, or equivalently for right moving spin up particles in the both bands and left moving spin down particles in both bands. A field of opposite helicity would have gapped out $\phi_{12c}+\theta_{12s}$ and $\phi_{21c}+\theta_{21s}$, corresponding to left moving spin up particles in both bands and right moving spin down particles in both bands.
 
 \section{RG flow due to the Overhauser fields}\label{append:first_flow}
Taking into account the Overhauser fields, the electrons are subject to the effective Hamiltonian
 
 \begin{align}\label{eq:ham_append_ov_u12}
H_{\rm e} &= \sum_{i=1,2}\sum_{j=c,s}\int \frac{dz}{2\pi}\,\left(\frac{u_{ij}}{K_{ij}}\,(\partial_z \phi_{ij})^2 + u_{ij} K_{ij} \, (\partial_z\theta_{ij})^2\right)\nonumber\\
 &+ \int \frac{dz}{2\pi}\,\frac{4U}{\pi}\, (\partial_z \phi_{1c})(\partial_z \phi_{2c})\\
  &+\sum_{i=1,2} \frac{B_{xy,i}}{2 \pi \alpha} \int dz\,\cos\left(\sqrt{2}(\phi_{ic} + \theta_{is})\right)~.\nonumber
\end{align}
 As discussed in Sec.~\ref{subsec:low_t}, the Overhauser fields $B_{xy,i}$ and the interband interaction $U$ have competing effects and, when taken on their own, drive the system to two different fixed points. Their combined effect is taken into account by first analyzing the flow due to the strongly relevant Overhauser field whilst neglecting the flow due to the marginal interband interactions, which will be treated in a second step.
 
At the initial fixed point described by the first line of Eq.~\eqref{eq:ham_append_ov_u12}, the cosine-terms of the Overhauser fields have the scaling dimensions $g_i = (K_{ic} + 1/K_{is})/2$. To integrate the flow of $B_{xy,i}$, we perform a canonical transformation that switches from spin and change degrees of freedom to fields $\phi_{i+} \sim \phi_{ic}+\theta_{is}$ whilst preserving the scaling dimension of the cosine-term with respect to the diagonal part of the Hamiltonian,

\begin{subequations}\label{eq:append_overhauser_trafo}
\begin{align}
 \phi_{ic} &=\frac{K_{ic}}{\sqrt{K_i}}\,\phi_{i+} + \sqrt{\frac{K_{ic}}{K_{is}\,K_i}}\,\phi_{i-}~,\\
  \theta_{ic} &=\frac{1}{\sqrt{K_i}}\,\theta_{i+} + \frac{1}{\sqrt{K_{ic}\,K_{is}\,K_i}}\,\theta_{i-}~,\\
  \phi_{is} &=\frac{1}{\sqrt{K_i}}\,\theta_{i+} - \sqrt{\frac{K_{is}\,K_{ic}}{K_i}}\,\theta_{i-} ~,\\
  \theta_{is}& = \frac{1}{K_{is}\sqrt{K_i}}\,\phi_{i+} - \sqrt{\frac{K_{ic}}{K_{is}\,K_i}}\,\phi_{i-}~,
\end{align}
\end{subequations}
with $K_i = K_{ic} + 1/K_{is} = 2g_{i}$. This transformation yields

\begin{align}
 H_{\rm e} &= H_{1} + H_2+H_{12}
\end{align}
with
\begin{align}
 H_{i} &= \int \frac{dz}{2\pi} \,\left(u_{i+} \left(\partial_z\phi_{i+}\right)^2 + u_{i+}\, \left(\partial_z\theta_{i+}\right)^2\right)\\
  &+\int \frac{dz}{2\pi} \,\left(u_{i-} \left(\partial_z\phi_{i-}\right)^2 + u_{i-}\,\left(\partial_z\theta_{i-}\right)^2\right)\nonumber\\
  &+\int \frac{dz}{2\pi} \,2 U_{i}\,\left((\partial_z\phi_{i+})\,(\partial_{z}\phi_{i-}) + (\partial_z \theta_{i+})\,(\partial_z \theta_{i-})\right)\nonumber\\
    &+ \frac{B_{xy,i}}{2 \pi \alpha} \int dz\,\cos\left(\sqrt{2K_{i}}\,\phi_{i+}\right)\nonumber\text{ ,}
\end{align}
where

\begin{subequations}
\begin{align}
 u_{i+} &= \frac{u_{ic} K_{ic}+ u_{is}/K_{is}}{K_i}~,\\
  u_{i-} &= \frac{u_{ic}/K_{is}+ u_{is}\,K_{ic}}{K_i}~,\\
  U_{i} &= \frac{u_{ic}-u_{is}}{K_i}\,\sqrt{\frac{K_{ic}}{K_{is}}} ~,
\end{align}
\end{subequations}
and where the marginal interband interaction reads

\begin{align}
H_{12} =  \int \frac{dz}{2\pi}\,&\frac{4U}{\pi}\, \left(\frac{K_{1c}}{\sqrt{K_1}}\,\partial_z \phi_{1+} + \sqrt{\frac{K_{1c}}{K_{1s}\,K_1}}\,\partial_z \phi_{1-}\right)\\
 &\times \left(\frac{K_{2c}}{\sqrt{K_2}}\,\partial_z \phi_{2+} + \sqrt{\frac{K_{2c}}{K_{2s}\,K_2}}\,\partial_z \phi_{2-}\right)~. \nonumber
\end{align}
From there, we derive the RG equation for $B_{xy,i}$ in a real space RG scheme.\cite{giamarchi_book} Parameterizing the running short distance cutoff as $\alpha(b) = b\,\alpha$, we obtain the flow of $B_{xy,i}$ as

\begin{align}
\frac{\partial B_{xy,i}}{\partial\log(b)} = (1-g_{i})\,B_{xy,i}~.
\end{align}

\section{Self-consistent derivation of the RKKY interaction}\label{append:self_cons_rkky}
In Appendix \ref{append:first_flow}, we established that the intermediate fixed point $c$ of Fig.~\ref{fig:rg_fixed_points}, reached after integrating the flow of the relevant Overhauser fields, can be described as 2 gapped and two gapless spinless Luttinger liquids. The remaining ungapped $\phi_{i-}$ modes now mediate an RKKY interaction similar to case discussed in Appendix \ref{Appendix:suscept}. We derive the latter in a mean field approximation where the $\phi_{i+}$ fields are locked to the minima of the cosine-terms. As one could expect, Gaussian fluctuations around this mean field have been shown to be unimportant,\cite{braunecker_prb_09} and will be neglected here. We also neglect renormalizations of the Luttinger liquid parameters and effective velocities during the flow between the fixed points $a$ and $c$ of Fig.~\ref{fig:rg_fixed_points} since the gaps $B_{xy,i}$ are strongly RG relevant and the associated flow is thus short. In addition, interband interactions are not important unless $v_{F1}\approx v_{F2}$, see main text. In this approximation, the mean field Hamiltonian of the remaining gapless $\phi_{i-}$ modes reads

\begin{align}
 H_{{\rm e},-} &= \sum_{i = 1,2}\int \frac{dz}{2\pi} \,\left(u_{i-} \left(\partial_z\phi_{i-}\right)^2 + u_{i-}\,\left(\partial_z\theta_{i-}\right)^2\right)\nonumber\\
  &+\int \frac{dz}{2\pi}\,\frac{4U_{\rm eff}}{\pi}\, \left(\partial_z \phi_{1-}\right)\,\left(\partial_z \phi_{2-}\right) ~,
\end{align}
where the effective interaction between the $\phi_{i-}$-modes is given by

\begin{align}
 U_{\rm eff} = U\, \sqrt{\frac{K_{1c}\,K_{2c}}{K_{1s}\,K_1\,K_{2s}\,K_2}} ~.
\end{align}
This Hamiltonian can be diagonalized with a transformation similar to Eq.~\eqref{eq:diagonalization_trafo}, namely

\begin{subequations}\label{eq:diagonalization_trafo_with_ov}
 \begin{align}
\phi_{1-}&= \sqrt{\frac{u_{1-}}{v_{+}\,(1+\mathcal{A}_-^2)}}\,\phi_{+}+\sqrt{\frac{\mathcal{A}_-^2\,u_{1-}}{v_{-}(1+\mathcal{A}_-^2)}}\,\phi_{-}~\\
\phi_{2-}&= \sqrt{\frac{\mathcal{A}_-^2\,u_{2-}}{v_{+}(1+\mathcal{A}_-^2)}}\,\phi_{+}- \sqrt{\frac{u_{2-}}{v_{-}(1+\mathcal{A}_-^2)}}\,\phi_{-}~,\\
\theta_{1-}&=\sqrt{\frac{v_{+}}{u_{1-}(1+\mathcal{A}_-^2)}}\,\theta_{+} + \sqrt{\frac{\mathcal{A}_-^2\,v_{-}}{u_{1-}(1+\mathcal{A}_-^2)}}\,\theta_{-}~, \\
\theta_{2-}&=\sqrt{\frac{\mathcal{A}_-^2\,v_{+}}{u_{2-}(1+\mathcal{A}_-^2)}}\,\theta_{+} - \sqrt{\frac{v_{-}}{u_{2-}(1+\mathcal{A}_-^2)}}\,\theta_{-} \,,
 \end{align}
\end{subequations}
with

\begin{align}
 v_{\pm} = \sqrt{\frac{u_{1-}^2+u_{2-}^2}{2}\pm\sqrt{\left(\frac{\delta u_{12-}^2}{2}\right)^2+\left(\frac{2U_{\rm eff}}{\pi}\right)^2 u_{1-} u_{2-}}}
\end{align}
where we used $\delta u_{12-}^2 = u_{1-}^2-u_{2-}^2$
and with
\begin{align}
 \mathcal{A}_-= \frac{(4U_{\rm eff}/\pi)\,\sqrt{u_{1-}u_{2-}}}{\sqrt{\left(\delta u_{12-}^2\right)^2+\left(4U_{\rm eff}/\pi\right)^2\,u_{1-} u_{2-}}+\delta u_{12-}^2} ~.
\end{align}
This transformation brings the gapless sector of the Hamiltonian in the diagonal form

\begin{align}
 H_{{\rm e},-} = \sum_{k=\pm}\frac{1}{2\pi} \int dz\,\left((\partial_z \phi_k)^2 + (\partial_z \theta_k)^2\right) ~,
\end{align}
which in turn allows us to evaluate the residual RKKY interactions due to the gapless electronic modes $\phi_{+}$ and $\phi_{-}$. As before, the residual RKKY interactions are determined by the residual spin susceptibility. We start from Eq.~\eqref{eq:append_chi}, apply the transformation \eqref{eq:append_overhauser_trafo} and drop the gapped fields $\phi_{i+}$ (which are simple constants in imaginary time and space and thus drop out). For the $x$ and $y$ components of the residual spin susceptibilities in the two bands, we obtain
\begin{align}
 \chi_{ii,q}^x &=  \chi_{ii,q}^y = \frac{-L}{4(2\pi\alpha)^2} \,\int d(z-z') e^{iq(z-z')}\,\Bigl(e^{-i2k_{Fi}(z-z')}\nonumber\\ 
 &\times\langle e^{i\sqrt{2}\sqrt{4K_{ic}/(K_{is} K_i)}(\phi_{i-}(z,\tau)-\phi_{i-}(z,\tau))}\rangle\\
 &+\rm{h.c.}\Bigr)~,\nonumber
\end{align}
and thus
\begin{align}
 \chi_{ii,q}^x& =  \chi_{ii,q}^y = \frac{-L}{4(2\pi\alpha)^2} \,\int d(z-z') e^{iq(z-z')}\,\Bigl(e^{-i2k_{Fi}(z-z')}\nonumber\\ 
 &\times\langle e^{i\sqrt{2}(A_i\phi_{-}(z,\tau)+B_i\phi_{+}(z,\tau)-A_i\phi_{-}(z,\tau)-B_i\phi_{+}(z,\tau))}\rangle\nonumber\\
 &+\rm{h.c.}\Bigr)~,
\end{align}
where
\begin{subequations}
\begin{align}
 A_1 &= \sqrt{\frac{4\,K_{1c}\,u_{1-}}{K_{1s}\, K_1\,v_{+}\,(1+\mathcal{A}_-^2)}}~,\\
B_1 &= \sqrt{\frac{4\,K_{1c}\,\mathcal{A}_-^2\,u_{1-}}{K_{1s}\, K_1\,v_{-}(1+\mathcal{A}_-^2)}}~,\\
 A_2 &= \sqrt{\frac{4\,K_{2c}\,\mathcal{A}_-^2\,u_{2-}}{K_{2s}\, K_2\,v_{+}\,(1+\mathcal{A}_-^2)}}~,\\
B_2 &= -\sqrt{\frac{4\,K_{2c}\,u_{2-}}{K_{2s}\, K_2\,v_{-}(1+\mathcal{A}_-^2)}}~.
\end{align}
\end{subequations}
Evaluating these expressions and neglecting the velocity difference between $\phi_+$ and $\phi_-$ yields the spin susceptibility as

\begin{align}
 &\left.\chi_{ii,q}^{x,\rm ret}\right|_{\omega\to0} =-\frac{L\,\sin\left(\pi g_{ii}'\right)}{16\pi^2 u_{i-}}\,\left(\frac{\beta u_{i-}}{2 \pi \alpha}\right)^{2-2g_{ii}'}\\
 &\times\sum_{\kappa=\pm}\Gamma^2\left(1-g_{ii}'\right)\,\left|\frac{\Gamma\left(\frac{g_{ii}'}{2}-i \frac{\beta u_{i-}}{4 \pi}(q+\kappa 2 k_{Fi})\right)}{\Gamma\left(\frac{2-g_{ii}'}{2}-i \frac{\beta u_{i-}}{4 \pi}(q+\kappa 2 k_{Fi})\right)}\right|^2\nonumber
\end{align}
where $\Gamma$ is the standard Gamma function and with

\begin{equation}
2g_{ii}'= A_i^2 + B_i^2~.
\end{equation}
An analogous calculation can also be performed to calculate $g_{12}'$. Due to the uncertainty principle, an ordering of $\phi_{i+}$ regularizes terms that depend on $\theta_{i+}$.\cite{starykh_gapped_99,braunecker_prb_12, meng_loss_spin_susceptibility_sll_13} Roughly speaking, an expectation value involving the fields $\theta_{i+}$ averages to zero. For the same reason, also the $z$ components of the susceptibilities are non-divergent. 

\section{Tyablikov decoupling and ordering temperatures}\label{append:tyablikov}
Neglecting the magnetic field due to the electron polarization, the symmetric sector of the  nuclear spins is governed by the Hamiltonian

\begin{align}
H_{\rm RKKY}' &= \frac{1}{N} \sum_{q,\alpha}I_{{\rm s},-q}^\alpha \,\frac{J_{{\rm s},q}^{\alpha\alpha}{}'}{N_\perp^2}\, I_{{\rm s},q}^\alpha\label{eq:H_rkky_3}~.
\end{align}
Using $I^\alpha_{{\rm s}, q} = \sum_q \,e^{-i q z_i}\, I^\alpha_{{\rm s}, i_{\parallel}}$ with $z_i = a \,i_{\parallel}$ and $(1/N)\,\sum_q e^{i q z_i}\, J_{{\rm s},q}^{\alpha\alpha}{}' = J_{{\rm s}}^{\alpha\alpha}{}'(z_i)$, we obtain the real space RKKY Hamiltonian
\begin{align}
H_{\rm RKKY}' &= \sum_{i,j,\alpha}I_{{\rm s},i_{\parallel}}^\alpha \,\frac{J_{{\rm s}}^{\alpha\alpha}{}'(z_i-z_j)}{N_\perp^2}\, I_{{\rm s},j_{\parallel}}^\alpha\label{eq:H_rkky_4}~.
\end{align}
This Hamiltonian allows us to calculate, for instance, the critical temperature $T^*_1$ for the ordering of the nuclear spins in a helix of momentum $2k_{F1}$ in the symmetric channel. It is at first useful to realize that the nuclear spin order as well as the RKKY interaction essentially only have Fourier components either close to $\pm2k_{F1}$ or $\pm2k_{F2}$. Since the different minima of the RKKY interaction do not overlap, these two sets of Fourier components can be analyzed independently. We can thus rewrite the Hamiltonian as

\begin{align}
H_{\rm RKKY}' &= \frac{1}{N} \sum_{q\approx \pm2k_{F1}}\sum_\alpha I_{{\rm s},-q}^\alpha \,\frac{J_{{\rm s},q}^{\alpha\alpha}{}'}{N_\perp^2}\, I_{{\rm s},q}^\alpha\\
&+\frac{1}{N} \sum_{q\approx \pm2k_{F1}}\sum_\alpha I_{{\rm s},-q}^\alpha \,\frac{J_{{\rm s},q}^{\alpha\alpha}{}'}{N_\perp^2}\, I_{{\rm s},q}^\alpha~,
\end{align}
and treat the Fourier modes close to $\pm k_{F1}$ and $\pm k_{F2}$ separately. Retaining at first only momenta $q\approx \pm2 k_{F1}$, we can go to a rotated frame of reference where this order corresponds to a ferromagnetic polarization. This is achieved by the transformation

\begin{align}
 \boldsymbol{I}_{{\rm s}, i_{\parallel}}^{\rm rot.} = \begin{pmatrix} \cos(2k_{F1}z_i)&\sin(2k_{F1}z_i)&0\\-\sin(2k_{F1}z_i)&\cos(2k_{F1}z_i)&0\\0&0&1\end{pmatrix} \boldsymbol{I}_{{\rm s}, i_{\parallel}}~,
\end{align}
which transforms the RKKY exchange coupling to

\begin{align}
 J_{\rm s}^{\rm rot.}{}'(z_i-z_j) = \begin{pmatrix}A_{ij}&B_{ij}&0\\B_{ji}&A_{ij}&0\\0&0&J_{\rm s}^{zz}{}'(z_i-z_j)\end{pmatrix}~.
\end{align}
with

\begin{subequations}
\begin{align}
 A_{ij} &= +A_{ji} =  J _{{\rm s}}^{xx}{}'(z_i-z_j)\,\cos(2 k_{F1} (z_i-zj))~,\\
 B_{ij} &= -B_{ji} = J _{{\rm s}}^{xx}{}'(z_i-z_j)\,\sin(2 k_{F1} (z_i-zj))~.
\end{align}
\end{subequations}
In this basis, the Overhauser field resulting from the Fourier components close to $2k_{F1}$ reads

\begin{align}\label{eq:actual_overhauser_field_trafo}
 \langle\boldsymbol{I}_{{\rm s},i_{\parallel}}^{\rm rot.}\rangle &=  I \, N_\perp \, m_{2k_{F1}}\begin{pmatrix}1\\0\\0\end{pmatrix} ~.
\end{align}
Next, we apply the so-called Tyablikov decoupling which is based on the decoupling of the equation of motion of the nuclear spin Green's function. The latter is defined as

\begin{align}
 G_{ij,+-}^{\rm ret}(t,t') = -i \theta(t-t')\,\langle\left[I_{{\rm s}, i_{\parallel}}^+(t), I_{{\rm s}, j_{\parallel}}^-(t')\right]\rangle
\end{align}
with $I_{{\rm s}, i_{\parallel}}^\pm(t) = I_{{\rm s}, i_{\parallel}}^y(t)\pm i  I_{{\rm s}, i_{\parallel}}^z(t)$. The equation of motion for this Green's function reads

\begin{align}
& i\partial_t \,G_{ij,+-}^{\rm ret}(t,t') =\delta(t-t')\,\langle\left[I_{{\rm s}, i_{\parallel}}^+(t), I_{{\rm s}, i_{\parallel}}^-(t')\right]\rangle\\
 &+i \theta(t-t')\,\langle\left[\left[H_{RKKY}',I_{{\rm s}, i_{\parallel}}^+(t)\right], I_{{\rm s}, j_{\parallel}}^-(t')\right]\rangle\nonumber~.
\end{align}
Fourier transformation of the equation of motion finally yields
\begin{align}
 &\int d(t-t')\,e^{i \omega (t-t')}\,\left(i\partial_t \,G_{ij,+-}^{\rm ret}(t,t')\right) \\
 &= \omega \,G_{ij,+-}^{\rm ret}(\omega)+\langle\left[I_{{\rm s}, i_{\parallel}}^+(t), I_{{\rm s}, i_{\parallel}}^-(t)\right]\rangle\nonumber\\
 &-\langle\langle\left[\left[H_{RKKY}',I_{{\rm s}, i_{\parallel}}^+\right], I_{{\rm s}, j_{\parallel}}^-\right]\rangle\rangle_{\omega}^{ret.}\nonumber
 \end{align}
 with
 
 \begin{align}
  &\langle\langle\left[\left[H_{RKKY}',I_{{\rm s}, i_{\parallel}}^+\right], I_{{\rm s}, j_{\parallel}}^-\right]\rangle\rangle_{\omega}^{\rm ret.} \\
  &= \int d(t-t')\,e^{i \omega (t-t')}\nonumber\\
 &\quad\quad\times -i\,\theta(t-t') \langle\left[\left[H_{RKKY}',I_{{\rm s}, i_{\parallel}}^+(t)\right], I_{{\rm s}, j_{\parallel}}^-(t')\right]\rangle\nonumber~.
 \end{align}
The commutation of the nuclear spin raising operator and the Hamiltonian can be evaluated as

\begin{align}
&\left[H_{RKKY}',I_{{\rm s}, i_{\parallel}}^+\right] = \sum_{m}\left\{\frac{A_{mi}}{N_{\perp}^2} \, (I_{{\rm s},m_{\parallel}}^x\,I_{{\rm s}, i_{\parallel}}^++I_{{\rm s}, i_{\parallel}}^+\,I_{{\rm s},m_{\parallel}}^x)\right.\\
&-\frac{A_{mi}+J_{{\rm s}}^{z}{}'(z_m-z_i)}{2N_{\perp}^2}\,(I_{{\rm s},i_{\parallel}}^x\,I_{{\rm s}, m_{\parallel}}^++I_{{\rm s}, m_{\parallel}}^+\,I_{{\rm s},i_{\parallel}}^x)\nonumber\\
&-\frac{A_{mi}-J_{{\rm s}}^{z}{}'(z_m-z_i)}{2N_{\perp}^2}\,(I_{{\rm s},i_{\parallel}}^x\,I_{{\rm s}, m_{\parallel}}^- + I_{{\rm s}, m_{\parallel}}^-\, I_{{\rm s},i_{\parallel}}^x)\nonumber\\
&+\frac{B_{mi}}{N_{\perp}^2} \, I_{{\rm s},m_{\parallel}}^+\,I_{{\rm s}, i_{\parallel}}^++\frac{B_{mi}}{2N_{\perp}^2}\,\left(I_{{\rm s},i_{\parallel}}^+\,I_{{\rm s}, m_{\parallel}}^-+I_{{\rm s},m_{\parallel}}^-\,I_{{\rm s}, i_{\parallel}}^+\right)\nonumber\\
&\left.+\frac{2 B_{mi}}{N_{\perp}^2}\,I_{{\rm s},m_{\parallel}}^x\,I_{{\rm s}, i_{\parallel}}^x\nonumber\right\} ~.
\end{align}
The essential approximation of the Tyablikov decoupling now consists of the following simplifications,
\begin{subequations}\label{eq:decoupling}
\begin{align}
\langle\langle \left[\left[I_{{\rm s},i_{\parallel}}^x, I_{{\rm s},j_{\parallel}}^\pm\right]_+ ,I_{{\rm s},m_{\parallel}}^-\right] \rangle\rangle_{\omega}^{\rm ret.}&=\\
2\langle I_{{\rm s},i_{\parallel}}^x \rangle\, & \langle\langle \left[ I_{{\rm s},j_{\parallel}}^\pm,I_{{\rm s},m_{\parallel}}^-\right] \rangle\rangle_{\omega}^{\rm ret.}\nonumber~,\\
\langle\langle \left[I_{{\rm s},i_{\parallel}}^x I_{{\rm s},j_{\parallel}}^x,I_{{\rm s},m_{\parallel}}^-\right] \rangle\rangle_{\omega}^{\rm ret.} &= 0~,\\
\langle\langle \left[I_{{\rm s},i_{\parallel}}^\pm I_{{\rm s},j_{\parallel}}^\pm,I_{{\rm s},m_{\parallel}}^-\right] \rangle\rangle_{\omega}^{\rm ret.} &= 0~,
\end{align}
\end{subequations}
where $[(\cdot),(\cdot)]_+$ is the anticommutator, while $[(\cdot),(\cdot)]$ denotes the anticommutator as before. At this point, these decouplings are mathematically not justified. They have however successfully been applied before and in this sense benefit from an a posteriori justification.\cite{nolting_book, Tyablikov_additional} Physically, the Tyablikov decouplings are indeed plausible. In the ordered state, the fluctuations around the ground state are small, such that one may approximate $I_{{\rm s},i_{\parallel}}^x \to \langle I_{{\rm s}}^x\rangle$ = \text{const.}, which implies the first and second line of Eq.~\eqref{eq:decoupling}. The third line may be interpreted as following from the fact that any decoupling would involve the ground state expectation value of a spin raising or lowering operator, which should, to good approximation, vanish in the semiclassical ground state. In addition, we also drop the purely local terms $\sim A_{ii}$, $J_{\rm s}^z(0)$. The equation of motion for the 
nuclear spin Green's function is then given by

\begin{align}
\tilde{\omega}\,G_{ij,+-}^{\rm ret.}(\tilde{\omega}) &= 2\delta_{ij}\, \langle I_{{\rm s}}^x\rangle\\
&-2\sum_m \frac{A_{mi}}{N_{\perp}^2}\, \langle I_{{\rm s}}^x\rangle\, G_{ij,+-}^{\rm ret.}(\tilde{\omega})\nonumber\\
&+\sum_m  \frac{A_{mi}+J_{{\rm s}}^{z}{}'(z_m-z_i)}{N_{\perp}^2}\, \langle I_{{\rm s}}^x\rangle\, G_{mj,+-}^{\rm ret.}(\tilde{\omega})\nonumber\\
 &+\sum_m\frac{A_{mi}-J_{{\rm s}}^{z}{}'(z_m-z_i)}{N_{\perp}^2}\, \langle I_{{\rm s}}^x\rangle\, G_{mj,--}^{\rm ret.}(\tilde{\omega})\nonumber~,
\end{align}
with $\tilde{\omega} = \omega + i\,0^+$. Now performing a second Fourier transformation to momentum space, defined as

\begin{align}
 G_{q,+-}^{\rm ret.}(\tilde{\omega}) = \frac{1}{N}\sum_{i,j}e^{iq(z_i-z_j)}\,G_{ij,+-}(\tilde{\omega})~,
\end{align}
yields

\begin{align}\label{eq:tyablikov_gpm}
\tilde{\omega}\,G_{+-}^{\rm ret.}&(q,\tilde{\omega}) = 2\, \langle I_{{\rm s}}^x\rangle\\
&-\langle I_{\rm s}^x\rangle \,\frac{2 A(q=0)-A(q)-J_{\rm s}^z(q)}{N_{\perp}^2}\,G_{+-}^{\rm ret.}(q,\tilde{\omega})\nonumber\\
&+\langle I_{\rm s}^x\rangle \,\frac{A(q)-J_{\rm s}^z(q)}{N_{\perp}^2}\,G_{--}^{\rm ret.}(q,\tilde{\omega})\nonumber~.
\end{align}
The Green's function $G_{mj,--}^{\rm ret.}(\tilde{\omega})$, i.e.~the Fourier transform of $G_{ij,--}^{\rm ret}(t,t') = -i \theta(t-t')\,\langle\left[I_{{\rm s}, i_{\parallel}}^-(t), I_{{\rm s}, j_{\parallel}}^-(t')\right]\rangle$, can similarly be shown to obey the equation of motion
\begin{align}\label{eq:tyablikov_gmm}
\tilde{\omega}\,G_{--}^{\rm ret.}&(q,\tilde{\omega}) = \\
&+\langle I_{\rm s}^x\rangle \,\frac{2 A(q=0)-A(q)-J_{\rm s}^z(q)}{N_{\perp}^2}\,G_{--}^{\rm ret.}(q,\tilde{\omega})\nonumber\\
&-\langle I_{\rm s}^x\rangle \,\frac{A(q)-J_{\rm s}^z(q)}{N_{\perp}^2}\,G_{+-}^{\rm ret.}(q,\tilde{\omega})\nonumber~.
\end{align}
We may thus define 

\begin{subequations}
 \begin{align}
  \tilde{\alpha}_q &= \langle I_{\rm s}^x\rangle \,\frac{2 A(q=0)-A(q)-J_{\rm s}^z(q)}{N_{\perp}^2}~,\\
\tilde{\beta}_q &= \langle I_{\rm s}^x\rangle \,\frac{A(q)-J_{\rm s}^z(q)}{N_{\perp}^2}~.
 \end{align}
\end{subequations}
and obtain by plugging Eq.~\eqref{eq:tyablikov_gpm} into Eq.~\eqref{eq:tyablikov_gmm} that

\begin{align}
 G_{+-}^{\rm ret.}(q,\tilde{\omega}) &= \frac{2\,\langle I_{\rm s}^x\rangle}{\tilde{\omega}+\sqrt{\tilde{\alpha}_q^2-\tilde{\beta}_q^2}} + \frac{2\,\langle I_{\rm s}^x\rangle\,\left(\tilde{\alpha}_q-\sqrt{\tilde{\alpha}_q^2-\tilde{\beta}_q^2}\right)}{\tilde{\omega}^2-\left(\tilde{\alpha}_q^2-\tilde{\beta}_q^2\right)} ~.
\end{align}
In a final step, we now have to self-consistently determine the magnetization $\langle I_{\rm s}^x\rangle$. To this end, we realize that quite generally the Green's function and the associated density fulfill

\begin{align}
 \langle I_{{\rm s},j}^- I_{{\rm s},i}^+ \rangle_{q} =  \int\, d\omega \, n_{B}(\omega)\,\left(-\frac{1}{\pi}\text{Im}\left\{G_{+-}^{\rm ret.}(q,\tilde{\omega})\right\}\right)~,
\end{align}
where $n_B$ is the Bose distribution function. From there, we find by Fourier transformation to real space that
\begin{align}
 \langle I_{{\rm s},i}^- I_{{\rm s},i}^+ \rangle &=\frac{2\,\langle I_{\rm s}^x\rangle}{N}\sum_q\left[\frac{\tilde{\alpha}_q}{\sqrt{\tilde{\alpha}_q^2-\tilde{\beta}_q^2}}\,n_B(\sqrt{\tilde{\alpha}_q^2-\tilde{\beta}_q^2})\right.\\
 &+\left.\frac{\tilde{\alpha}_q-\sqrt{\tilde{\alpha}_q^2-\tilde{\beta}_q^2}}{2\sqrt{\tilde{\alpha}_q^2-\tilde{\beta}_q^2}}\right]\nonumber
\end{align}
At the same time, we know from the angular momentum algebra that $I_{{\rm s},i}^- I_{{\rm s},i}^+ = (IN_{\perp})(IN_{\perp}+1) - I_{{\rm s},i}^x-\left(I_{{\rm s},i}^x\right)^2$. This leads us to a self-consistent equation for $\langle I_{\rm s}^x\rangle$ (through $\tilde{\alpha}_q$ and $\tilde{\beta}_q$),

\begin{align}
 \langle I_{\rm s}^x\rangle &= \frac{p\,I N_{\perp}\left(I N_{\perp}+1\right)}{1+\frac{a}{\pi}\int dq\,\frac{\tilde{\alpha}_q}{\sqrt{\tilde{\alpha}_q^2-\tilde{\beta}_q^2}}\,n_B(\sqrt{\tilde{\alpha}_q^2-\tilde{\beta}_q^2})+\frac{\tilde{\alpha}_q-\sqrt{\tilde{\alpha}_q^2-\tilde{\beta}_q^2}}{2\sqrt{\tilde{\alpha}_q^2-\tilde{\beta}_q^2}}}\text{ ,}
\end{align}
where $p\,I N_{\perp}\left(I N_{\perp}+1\right) = I N_{\perp}\left(I N_{\perp}+1\right)-\langle \left(I_{{\rm s},i}^x\right)^2\rangle$. Due to the appearance of $\langle \left(I_{{\rm s},i}^x\right)^2\rangle$, this equation is not sufficient to determine the magnetization for $IN_\perp > 1/2$. One rather has to construct a set of $2IN_{\perp}-1$ coupled equations by evaluating Green's functions of the form $G^n \sim \langle \left[I_{\rm s}^+,(I_{\rm s}^z)^n\,I_{\rm s}^-\right] \rangle$, see Ref.~[\onlinecite{nolting_book}]. To determine the critical temperature, however, one may recall that for $T = T_{c1}$, the magnetization $\langle I_{\rm s}^x\rangle$ considered here (i.e. the Fourier components close to $q = 2k_{F1}$) vanishes. For $T\approx T_{c1}$, the expectation value $\langle\left(I_{{\rm s},i}^x\right)^2\rangle$ is dominated by the large fluctuations of the magnetization around the ordered state. It is thus given by $\langle\left(I_{{\rm s},i}^x\right)^2\rangle \approx I N_{\perp}(I N_{\perp}+1)/3$, since the system is disordered for $T = T_{c1}^+$, and spin rotation symmetry is not broken at high temperatures. Because we are considering a finite size wire, the average $\langle\left(I_{{\rm s},i}^x\right)^2\rangle$ is also not allowed to jump at the transition. This sets $p \approx 2/3$. With this approximation, we can finally expand the Bose function for $T = T_{c1} - \delta T$ in the small magnetization $\langle I_{\rm s}^x\rangle$ and solve the self-consistent equation. Approximating finally $IN_{\perp}(IN_{\perp}+1) \to (IN_{\perp})^2$ yields

\begin{align}
 T_{c1} = \frac{2\,I^2}{3}\,\left(\frac{a}{\pi}\,\int dq\,\frac{\tilde{\alpha}'(q)}{\tilde{\alpha}'(q)^2-\tilde{\beta}'(q)^2}\right)^{-1}~,
\end{align}
with
\begin{subequations}
 \begin{align}
  \tilde{\alpha}'(q) &= \left[2 A(q=0)-A(q)-J_{\rm s}^z(q)\right]~,\\
\tilde{\beta}'(q) &= \left[A(q)-J_{\rm s}^z(q)\right]~.
 \end{align}
\end{subequations}

The critical temperature of the second band can be derived analogously.



\begin{thebibliography}{99}
\bibitem{braunecker_magnon_comment}
Note that the magnon analysis as given in Ref.~[\onlinecite{braunecker_prb_09}] misses the correct diagonalization of the magnon spectrum using a Bogoliubov transformation that preserves the bosonic commutation relations. The correct calculation however comes to the same conclusions.
\end{thebibliography}
\end{document}